\newcommand{\lya}{\mbox{Ly$\alpha$}}

\newcommand{\halpha}{\mbox{H$\alpha$}}

\newcommand{\hI}{\mbox{H~{\sc i}}}
\newcommand{\hII}{\mbox{H~{\sc ii}}}
\newcommand{\heII}{\mbox{He~{\sc ii}}}

\newcommand{\oII}{\mbox{O~{\sc ii}}}
\newcommand{\oIII}{\mbox{O~{\sc iii}}}

\newcommand{\cII}{\mbox{C~{\sc ii}}}
\newcommand{\cIII}{\mbox{C~{\sc iii}}}
\newcommand{\cIV}{\mbox{C~{\sc iv}}}

\newcommand{\kms}{\mbox{km\,s$^{-1}$}}
\newcommand{\ergsec}{\mbox{erg~s$^{-1}$}}

\newcommand{\ergseccmaa}{\mbox{erg~s$^{-1}$~cm$^{-2}$~\AA$^{-1}$}}

\newcommand{\msun}{\mbox{M$_\odot$}}

\newcommand{\msunyr}{\mbox{M$_\odot$~yr$^{-1}$}}
\newcommand{\fesclyc}{\mbox{$f_\mathrm{esc}^\mathrm{LyC}$}}

\newcommand{\ebv}{\mbox{$E_{B-V}$}}

\newcommand{\nhi}{\mbox{$N_\mathrm{HI}$}}
\newcommand{\llya}{\mbox{$L_{\mathrm{Ly}\alpha}$}}

\newcommand{\ewlya}{\mbox{$W_{\mathrm{Ly}\alpha}$}}

\newcommand{\tauigm}{\mbox{$\tau_\mathrm{IGM}$}}
\newcommand{\tauigmlya}{\mbox{$\tau_\mathrm{IGM}^{\mathrm{blue}}$}}
\newcommand{\zsys}{\mbox{$z_\mathrm{sys}$}}
\newcommand{\Lbluered}{\mbox{$L_\mathrm{B/R}$}}

\documentclass[a4paper,fleqn,usenatbib,twocolumn]{aastex63}

\usepackage[T1]{fontenc}
\usepackage{ae,aecompl}

\usepackage{graphicx}	% Including figure files
\usepackage{amsmath}	% Advanced maths commands
\usepackage{amssymb}	% Extra maths symbols

\received{June 3, 2020}
\revised{December 6, 2020}
\accepted{\today}
\shorttitle{Lyman alpha profiles and the IGM}
\shortauthors{M. J. Hayes et al.}

\begin{document}

\title{Spectral shapes of the Lyman-alpha emission from galaxies: I. blueshifted emission and intrinsic invariance with
redshift \footnote{Based on observations made with the NASA/ESA Hubble Space
Telescope, obtained from the data archive at the Space Telescope Science
Institute. STScI is operated by the Association of Universities for Research in
Astronomy, Inc. under NASA contract NAS 5-26555.}}

\correspondingauthor{Matthew J. Hayes}
\email{matthew@astro.su.se}

\author[0000-0001-8587-218X]{Matthew J. Hayes}
\affiliation{Stockholm University, Department of Astronomy and Oskar Klein
Centre for Cosmoparticle Physics, AlbaNova University Centre, SE-10691,
Stockholm, Sweden.}

\author[0000-0002-1025-7569]{Axel Runnholm}
\affiliation{Stockholm University, Department of Astronomy and Oskar Klein
Centre for Cosmoparticle Physics, AlbaNova University Centre, SE-10691,
Stockholm, Sweden.}

\author[0000-0003-2491-060X]{Max Gronke}
\affiliation{Department of Physics \& Astronomy, Johns Hopkins University,
Baltimore, MD 21218, USA and Hubble fellow}

\author[0000-0002-9136-8876]{Claudia Scarlata}
\affiliation{Minnesota Institute for Astrophysics, School of Physics and
Astronomy, University of Minnesota, 316 Church Str. SE, Minneapolis,MN 55455,
USA}

\begin{abstract} 
We demonstrate the redshift-evolution of the spectral profile of \hI\
Lyman-alpha (\lya) emission from star-forming galaxies.  In this first study we
pay special attention to the contribution of blueshifted emission.  At redshift
$z= 2.9-6.6$, we compile spectra of a sample of 229 \lya-selected galaxies
identified with the Multi-Unit Spectroscopic Explorer at the Very Large
Telescope, while at low-$z$ ($<0.44$) we use a sample of 74
ultraviolet-selected galaxies observed with the Cosmic Origin Spectrograph
onboard the Hubble Space Telescope.  At low-$z$, where absorption from the
intergalactic medium (IGM) is negligible, we show that the ratio of
\lya\ luminosity bluewards and redwards of line center (\Lbluered) increases
rapidly with increasing equivalent width (\ewlya). This correlation
does not, however, emerge at $z=3-4$, and we use bootstrap simulations to
demonstrate that trends in \Lbluered\ should be suppressed by
variations in IGM absorption.  Our main result is that the observed
blueshifted contribution evolves rapidly downwards with increasing redshift:
\Lbluered\ $\approx 30$~\% at $z\approx 0$, but drops to 15~\% at $z\approx 3$,
and to below 3~\% by $z\approx 6$.  Applying further simulations of the IGM
absorption to the unabsorbed COS spectrum, we demonstrate  that this decrease
in the blue-wing contribution can be entirely attributed to the thickening of
intervening \lya\ absorbing systems, with no need for additional \hI\ opacity
from local structure, companion galaxies, or cosmic infall.  We discuss our
results in light of the numerical radiative transfer simulations, the evolving
total \lya\ and ionizing output of galaxies, and the utility of
resolved \lya\ spectra in the reionization epoch.
\end{abstract}


\section{Introduction}\label{sect:intro}

The \lya\ emission line, and its spectral profile, encode a wealth of
information about both star-formation and various gas phases in galaxies,
including the interstellar medium (ISM), the more tenuous circumgalactic medium
(CGM) and, at the highest redshifts, the intergalactic medium (IGM).  Because
of the abundance of atomic hydrogen and the nature of the \hI\ atom, \lya\ is
intrinsically the brightest spectral feature of ionized nebulae, but also has
the largest optical depth to absorption.  \lya\ is therefore sufficiently
luminous to be observed from redshifts ($z$) out to above 9
\citep[e.g.][]{Hashimoto.2018}, but is then multiply scattered by \hI\ atoms,
and re-shaped by radiative transfer (RT) effects into which many galaxy
properties enter (see \citealt{Runnholm.2020lars} for a thorough investigation,
and \citealt{Dijkstra.2014} and \citealt{Hayes.2015} for reviews).

This absorption by \hI\ may occur in both `conventional' (=\hI-dominated)
atomic gas clouds or residual \hI\ ions in \hII-dominated gas.  The \lya\ RT,
therefore, can be a problem covering various regions and physical scales.  The
emergent \lya\ line profile will firstly be a reflection of the kinematics of
the \hII\ regions in which \lya\ is intrinsically produced.  However this
velocity profile is subsequently modulated by scattering events within the
\hII\ region itself, that will introduce some degree of broadening depending
upon the small scale gas densities.  But \lya\ photons may also have to
traverse high column density, optically thick gas, that permits escape only
after long excursions in frequency space \citep[wing
scatterings][]{Adams.1972,Neufeld.1990}.  This generally high optical depth at
line center introduces double-peaked profiles, consisting of blueshifted and
redshifted components with little emission at line center ($\Delta v=0$), and a
separation that depends on the \hI\ column density (\nhi).  Because gas is
rarely static or homogeneous in the ISM, the density distribution, clumpiness,
and dynamics of the neutral and dense ionized medium in galaxies will all
influence the line profile.  As starburst galaxies typically show
moderate-velocity outflowing media (see reviews by \citealt{Rupke.2018} and
\citealt{Veilleux.2020}  at low-$z$; \citealt{Erb.2015} at high-$z$) stronger
redshifted \lya\ peaks, with secondary weaker blue components, are commonplace. 

Because the blueshifted component of \lya\ is easily suppressed by scattering
in outflowing material, significant emission bluewards of the \lya\ line center
($\Delta v<0$) is relatively rare, and has become a key signpost of a low gas
column density.  \citet{Henry.2015} first obtained UV spectra of a sample of
`Green Pea' galaxies: not only do these galaxies exhibit among the strongest
\lya\ lines known (average \lya\ escape fraction of $\approx 25$\% and EW of
$\approx 60$~\AA), but a remarkable fraction of 9/10 also show double peaks.
Similar results have since been presented in other low-$z$ samples with
similarly strong nebular EWs, and also high [\oIII]/[\oII] ratios
\citep{Yang.2017prof,Jaskot.2017,Izotov.2018,Izotov.2020}.  In Lyman break
galaxies (LBG) at $z=2-3$, the relative contribution of blueshifted \lya\
(defined as the ratio of equivalent widths at velocities bluewards and
redwards of line center) also correlates strongly with the \lya\ EW
\citep{Erb.2014}, and becomes even more prominent in very highly ionized
galaxies \citep{Erb.2016, Trainor.2016}.  These associations are also borne out
by theoretical calculations and models.  \citet{Verhamme.2015} used \lya\
radiation transfer simulations to show that a narrow velocity separation
between two spectrally resolved peaks could be used as a signifier of column
densities close to the limit of optical thickness at the Lyman edge; see
\citet{Osterbrock.1962} and \citet{Hummer.1971} for the original discussions in
the low optical depth limits.  The implication that galaxies with narrow \lya\
separations of below $\approx 250$~\kms\  would be Lyman continuum (LyC)
emitting galaxies was qualitatively borne out by recent observations (e.g.
\citealt{Izotov.2018}, see also \citealt{Jaskot.2019}) and further theoretical
work \citep[e.g.][]{Kimm.2019,Kakiichi.2019}.  As optical emission lines are
hard to detect with high fidelity at $z\gtrsim 6$ because of redshifting to the
IR, and absorption lines being equally hard because of the faint continuum, the
\lya\ spectral profile is a promising diagnostic for LyC emission at high-$z$.

Because \lya\ is absorbed by \hI, and the ionized fraction of the universe
evolves with redshift, \lya\ is a natural probe of the epoch of reionization
(EoR; most recently reviewed by \citealt{Dijkstra.2014}).  Studies began using
first-order tests of the evolving \lya\ luminosity function (LF; e.g.
\citealt{Kashikawa.2006,Malhotra.2004,Malhotra.2006}) but, in light of the fact
that galaxies may also assemble rapidly near the EoR, moved on to differential
tests of \lya\ evolution with respect to the underlying galaxy population.
This has been cast as either the `volumetric' \lya\ escape fraction
\citep{Hayes.2011evol,Dijkstra.2013,Konno.2016,Wold.2017} or as the fraction of
strong LAEs among LBGs
\citep{Stark.2010,Ono.2012,Curtis-Lake.2012,Schenker.2012,Pentericci.2014,deBarros.2017,Mason.2018boost,Kusakabe.2020}.

While the evolution of the average \lya\ output may probe the neutral gas
content of the universe at a certain redshift, reionization was also
inherently patchy because of the clustering of both cosmic gas and ionizing
sources.  Thus we need to ultimately progress beyond single averages at a given
cosmic time (as in the studies above).  Given that the EoR history and topology
is a function of both structural over-density (including gas) and galaxy
formation, it naturally follows that the blunt measure of neutral fraction vs
time will vary with position (analogous to downsizing).  Ultimately it must
give way to a more nuanced approach, accounting for spatial/environmental, as
well as temporal, variation.  \lya\ spectroscopy most likely offers the
solution here, but in the form of a more detailed, spectrally resolved approach
that targets the line profile.  

To kick-start this science, a number of double-peaked \lya\ emission lines have
recently been identified at high-$z$.  These include Aerith B at $z=5.8$
\citep{Bosman.2020}, NEPLA4 at $z=6.5$ \citep{Songaila.2018} and COLA1 at
$z=6.6$ \citep{Hu.2016}, which all show impressive \lya\ double-peaks, strongly
resemblant of the $z\approx 0$ compact starbursts
\citep{Henry.2015,Jaskot.2017,Izotov.2020}.  According to simplified,
homogeneous prescriptions for the neutral fraction of the universe at these
redshifts, these profiles should not exist, and even in modern radiation
hydrodynamic simulations of the EoR, blue peaks at $z\gtrsim 6$ are rare
\citep{Gronke.2020}.  Clearly the reionization progression is patchy, and
Aerith B is a perfect example, being discovered inside the proximity zone of a
luminous quasar.  \citet{Bagley.2017} recently calculated the bubble sizes
necessary to explain the existence of two very luminous \lya-emitter at
$z\approx 6.4$, finding they must exist on Mpc scales.  \citet{Matthee.2018}
performed a similar study using the spectral profile of COLA1 including its
blue \lya\ bump, and found comparable if somewhat smaller values; their
calculation was revised upwards very recently by \citet{MasonGronke.2020} whose
Bayesian inference places additional new constraints on the ionizing output and
neutral fraction within the \hII\ bubble, as well as the bubble radius itself.  

The uncertainties and degeneracies in these approaches are significant, but
will doubtless improve as more systems are discovered and data improve.  While
the universe beyond $z\approx 5$ is really the point of interest, these studies
can be significantly informed by those from low- and mid-$z$.  Specifically
with regard to the \lya\ profile analysis, it is never clear at high-$z$ how
much absorption can be attributed to the IGM, and how much is simply frequency
redistribution in the ISM/CGM.  Because galaxy populations evolve in many ways
-- mass, morphology, dust content, etc -- we do not believe a priori that \lya\
profiles from low-$z$ should necessarily hold in a galaxy population that is 10
times younger/less evolved. However, it remains well-motivated to see how the
high-$z$ observables would look in the no-evolution case.  We do not currently
have the data to probe these effects deep into the EoR, but are able to test
the evolution of the \lya\ spectral profile out to $z \gtrsim  6$ with existing
large samples, which is very much the first step.

In this paper we study the evolution of the \lya\ spectral line shape with
redshift, and the impact of absorption from discrete \hI-absorbing systems in
the IGM.  We use large samples of homogeneously-selected \lya\ spectra that
have recently become available: at high-$z$ we adopt the MUSE-WIDE data
(\citealt{Urrutia.2019}; see also \citealt{Herenz.2017musewide}) in the
CANDELS-Deep region of the GOODS-South field.  These data provide spectra with
resolving powers of $R(\equiv\lambda/\Delta\lambda)=1800$-3500 over 44
arcmin$^2$, of 478 galaxies.  For a baseline \lya\ spectrum that is not
attenuated by the IGM, we leverage various samples of low-$z$ spectra observed
with HST/COS.  These observations are pointed at pre-selected galaxies, but
have higher spectral resolving power ($R$ up to 22,000) and are also obtained
at redshifts where the IGM has little or no impact on the \lya\ spectral
profile. 

We present an overview of the data and samples in Section~\ref{sect:data}, and
describe the measurements we make and methods we use in
Section~\ref{sect:methods}, which mainly relate to estimating systemic
redshifts and simulating the IGM.  In Section~\ref{sect:sample} we present an
overview of the basic \lya\ properties of the high- and low-$z$ samples. 
We describe the shapes of the ensemble \lya\
profiles as a function of other key \lya\ observables in
Section~\ref{sect:res:stack}.  We present our results concerning redshift
evolution in Section~\ref{sect:res:zevol}, where we mainly show that low-$z$
spectral profiles, and a simply simulated IGM, can entirely reproduce the
evolution of the \lya\ spectral shape out to $z=6$.  In
Section~\ref{sect:discussion} we discuss these results in light of existing
\lya\ radiation transfer simulations and current knowledge of the IGM opacity.
Section~\ref{sect:outlook} presents some thoughts on the future application and
limitations of our approach, and we present a final summary of our findings in
Section~\ref{sect:conclusions}.  We adopt cosmological parameters of
$\{H_0,~\Omega_\mathrm{M},~\Omega_\Lambda \} =
\{70~\mathrm{km~s^{-1}~Mpc^{-1}},~0.3,~0.7\}$.

\section{Observational Data}\label{sect:data}

For this paper we draw upon a large compilation of data.  We use
HST/COS for low-$z$ spectroscopy of the \lya\ line, and VLT/MUSE for \lya\
spectroscopy at high-$z$.  As the UV continuum of the high-$z$ sample is
relatively faint, we use deep imaging from HST/ACS to determine the UV
magnitudes of the \lya-emitting galaxies at $z>2.9$.  We now describe the
origin, assembly, and processing of these data in turn.

\subsection{The low-z reference data: HST/COS}\label{sect:data:cos}

\subsubsection{Sample Overview}

HST/COS \citep{Green.2012} is an ultraviolet spectrograph onboard HST and in
the settings we use provides spectra of moderate nominal resolution,
with $R \approx$ 18,000 on average, for point-sources.  It has a 2\farcs5
diameter entrance aperture, which naturally will sample a range of
spatial scales, depending upon the redshift of the galaxy, which varies from
object to object.  Galaxies observed with HST/COS have all been pre-selected by
the HST observers and identified as the best targets with a certain scientific
objective in mind.

The COS data are drawn from the following general observer programs (with
principal investigators in parentheses): GO\,11522 (PI: Green), 11727
(Heckman), 12027 (Green), 12269 (Scarlata), 12583 (Hayes), 12928 (Henry), 13017
(Heckman), 13293 (Jaskot), 13744 (Thuan), 14080 (Jaskot), 14201 (Malhotra),
14635 (Izotov), 15136 (Izotov).  Most of the spectra have been published in
studies of galaxy winds and outflows, as well as their relation to \lya\ output
and kinematic properties \citep[][and
more]{Heckman.2011,Heckman.2015,Wofford.2013,Rivera-Thorsen.2015,Rivera-Thorsen.2017,Henry.2015,Jaskot.2014,Jaskot.2017,Jaskot.2019,Izotov.2016bLyc,Izotov.2018,Yang.2017prof}.
It is beyond the scope of this paper to summarize the results of these studies
but the selection identifies galaxies in the redshift range between $z=0.020$
and $z=0.44$, with a median value of $z=0.177$.  This places the \lya\ emission
line in either the G130M or G160M grating of COS, and delivers a spectral
resolution for point-sources between $R=$13,000 and 22,000 depending
upon lifetime position and redshift.  This resolution, however, depends
on spatial extent of the source compared to the aperture, which for \lya\ is
uncertain and the actual resolution is unknown in practice. We discard
low-resolution spectra obtained with the G140L grating, which has also been
obtained in some of the programs for a handful galaxies.  

The large majority of the targets were selected by UV emission,  usually from
GALEX far-UV photometry but also slit-less spectroscopy of \lya\ in the case of
GO\,12269.  Most were also constrained in at least redshift by optical line
emission, which almost exclusively comes from the Sloan Digital Sky Survey
(SDSS).  Further constraints may have been applied in terms of optical
compactness, \halpha\ or [\oIII] equivalent widths, [\oIII]/[\oII] ratios, and
more.  Attending mainly to the papers above, star formation rates (SFR) lie in
the range from 0.1 to over 50~\msunyr.

\subsubsection{Reduction and basic processing}

All spectra were obtained from the Mikulski Archive for Space Telescopes
(MAST), and reduced homogeneously with the calibration pipeline (CALCOS),
v.3.3.7.  Systemic redshifts have been derived from simultaneously fitting up
to 20 optical emission lines for sources with SDSS spectra, and from literature
values \citep[e.g.][]{Cowie.2011}, otherwise. More details can be found in
\citet{Runnholm.2020lasd}.

We first correct the COS spectra for Milky Way foreground extinction,
by looking up the mean \ebv\ at the target coordinates from the maps of
\citet[][using the \texttt{irsa\_dust} interface in
\texttt{astroquery}]{Schlafly.2011} and applying the \citet{Cardelli.1989}
extinction law.  No correction for internal extinction is made. We
continuum-subtract the COS spectra by fitting low-order polynomial functions to
the continuum over the wavelength range of $\lambda> 1219$~\AA, and
avoiding strong stellar and interstellar features.  The COS sample was mostly
UV-selected as discussed above, and the galaxies are not guaranteed to
have \lya\ in emission; as \lya\ emission is mandatory for this study we
discard all galaxies with signal-to-noise ratio in \lya\ less than 8.  This
leaves a sub-sample of 74 galaxies, for which the median \lya\ luminosity is
$1.25\times10^{42}$~\ergsec.

\subsection{High-z \lya\ information from MUSE}\label{sect:data:muse}

\subsubsection{Sample Overview}

Unlike HST/COS, VLT/MUSE \citep{Bacon.2010,Bacon.2014} is a survey instrument.
It is the first truly large-format integral field spectrograph, with a
field-of-view of $60\times 60$~arcsec$^2$.  MUSE therefore delivers an
emission line survey for \lya-emitting galaxies between $z=2.9$ and $z=6.6$
with 100~\% spectroscopic completeness.  However the spectral resolution is
somewhat lower than that of COS, and varies between $R = 1800$ at the blue end,
and 3500 at the red end. 

Many \lya\ surveys have already been conducted with MUSE, but here we rely upon
the MUSE-WIDE survey \citep{Herenz.2017musewide,Urrutia.2019}.  The main
reasons is that there are many pointings (44 fields), that are observed to a
homogeneous depth (1-hour per pointing).  The 1-dimensional data are easily
accessible through the CDS/VisieR database.  

\subsubsection{Basic processing}

MUSE-WIDE LAEs are continuum-subtracted using the same polynomial-fitting
method as the COS spectra (Section~\ref{sect:data:cos}).  The \lya\ luminosity
is measured by numerical integration over a window that is 2500~\kms\ wide, and
the errors are estimated by end-to-end Monte Carlo simulations.  As with the
COS spectra, we retain only spectra for which the SNR in \lya\ exceeds 8.  This
leaves a sub-sample of 229 galaxies, for which the median \lya\ luminosity is
$3.1\times10^{42}$~\ergsec.

We examined the MUSE-WIDE webpages to determine the sky conditions and
data quality when the data were obtained.  Using the SQL query
interface\footnote{\href{https://musewide.aip.de/query/}{https://musewide.aip.de/query/}}
we determine that 42 of the 44 fields were observed under either photometric or
clear conditions, implying the large majority of the LAEs in our sample should
have good photometric accuracy.  The MUSE-WIDE team has contrasted synthetic
photometry of the Data Release 1 spectra with HST magnitudes of brighter stars,
finding generally good agreement: the final dispersion in flux is on the order
of 5~\%, with all but a handful of sources agreeing to better than 10~\% (T.
Urrutia \& L. Wisotzki, private communication).  We also note that the main
results of this paper are based upon normalized spectra, for which absolute
photometric accuracy is not vital.

We finally note that this \lya\ selection at high-$z$, and UV selection at
low-$z$, are very different strategies for identifying galaxies.  This may
introduce some bias, but is a necessary condition of contrasting low-$z$
samples with large datasets at high-$z$ with current facilities.  We later
perform an additional level of matching between the two samples; see
Section~\ref{sect:samplematch} for more details.

\subsection{High-z continuum information from HST/ACS}\label{sect:data:acs}

Because the high-$z$ targets are selected based upon flux in the \lya\ line, a
subset of the sample will be faint or undetected in the continuum when
examining the same spectroscopic data.  The consequence is that the EWs
measured from the spectra themselves (see Section~\ref{sect:measquant}) will
become more imprecise towards faint magnitudes, and instead we adopt the UV
luminosities from HST broadband imaging.  The objects in the
\citet{Urrutia.2019} data release contain identifiers to the HST-identified
objects in the catalogs of \citet{Guo.2013} and \citet{Skelton.2014}.  This
catalog matching is an involved process, of which many details can be found in
Urrutia et al.  In short, photometric counterparts were determined by searching
for sources with 0.5~arcsec of the MUSE coordinates, followed by extensive
visual inspection to remove foreground sources, and comparison with existing
spectroscopic and photometric redshifts. We search the \citet{Guo.2013} for
photometry in the five ACS bands, F435W, F606W, F775W, F814W, and F850LP.
Naturally, the MUSE and HST photometry are extracted from different spatial
scales: in the case of MUSE data this is 3 Kron radii with a lower limit of
0.6~arcsec, while in the case of HST catalogs are AUTO magnitudes (minimum size
of 0\farcs125) with aperture corrections applied.  For additional information
concerning object deblending, we refer the reader to \citet{Guo.2013}.  Thus
the MUSE and HST photometry are not extracted from the same physical regions of
the data, but nor should they be: these available estimates are likely the best
global measurements of \lya\ and UV continuum that can be attained.

For the redshift of each galaxy, we first determine which is the bluest filter
in the ACS set that has a central wavelength redwards of \lya, so that it
samples more the stellar continuum than the IGM absorption.  Then we calculate
the contribution of the \lya\ line (flux from \citealt{Urrutia.2019}) to the
broadband magnitude, allowing for both the filter width and relative throughput
at the specific wavelength of \lya.  We subtract this value from the broadband
flux, and adopt the final line-subtracted magnitude for the galaxies' stellar
continuum.  

Other UV emission lines, such as \heII$\lambda 1640$,
\cIV$\lambda\lambda1548,1551$, and \oIII]$\lambda\lambda 1661,1666$ may fall
within the HST bandpasses, and may in principle contribute to the inferred
continuum flux. We note, however, that in the Urrutia et al sample only two
objects have detections of other UV lines.  Even at fainter magnitudes only a
few percent of \lya-selected objects have detections in the
\cIII]$\lambda\lambda 1907,1909$ doublet (17 of 692 galaxies;
\citealt{Maseda.2017} c.f. \citealt{Inami.2017}).  The potential contaminants
(\heII, \cIV, and \oIII]) may all have EWs comparable to \cIII], which in the
sample of \citet{Maseda.2017} has a mean value of 7.3~\AA\ with a standard
deviation of 5.1~\AA.  At $z\sim 3$ the observed EW would be $\sim 30$~\AA, and
would contribute only $\sim 1$~\% of the flux in the broad ACS/F606W filter.
In a full sample (including non-detections) the average EW will be much lower,
and in the more luminous (presumably more metallic) MUSE-WIDE sample, the
contribution will be lower still.  We therefore expect a negligible
contamination of the \lya\ EWs from other UV emission lines.

\section{Methods and Measurements}\label{sect:methods}

Here we describe the methods we use to make measurements in the \lya\ spectra
(Section~\ref{sect:measquant}), estimate the systemic redshifts
(Section~\ref{sect:zsys}), approximately match the spectral resolution between
COS and MUSE (Section~\ref{sect:specres}), and simulate the effects of
absorption in the IGM (Section~\ref{sect:abssim}). Details of the stacking
procedure can be found where they are first applied, in
Section~\ref{sect:res:stack}. 

\subsection{Measurements and quantities}\label{sect:measquant}

We adopt a modified version of the software implemented within the \emph{Lyman
alpha Spectral Database} (LASD\footnote{\url{http://lasd.lyman-alpha.com}};
\citealt{Runnholm.2020lasd}).  The LASD measures 38 spectroscopic/kinematic
properties of the \lya\ spectra, most of which are fluxes/flux densities or
velocities.  In this study we also use \lya\ luminosities derived from the
LASD, which are computed by numerical integration over a spectral region
2500~\kms\ wide around \lya.   This is done after continuum subtraction, and
the continuum estimate is used to derive the equivalent width, \ewlya.
The ratio of luminosities bluewards and redwards of line center
(\Lbluered) is a key quantity in this article, and is calculated in an
analogous way as the total luminosity: by numerical integration over the ranges
$-1250- 0$~\kms\ (blue) and $0-1250$~\kms\ (red).

As \lya\ is a resonance line it can be absorbed from the stellar continuum as
well as produced in emission by nebulae/\hII\ regions; in this study 
%we define the \lya\ flux as the total integral over the velocity range with 
we make no correction for blueshifted absorption.  This is in part motivated by
the fact that ostensibly pure emission lines can still be attenuated to an
unknown degree, and in part because the high-$z$ galaxies are too faint to
detect the continuum and the comparable correction could not be applied.  More
discussion on this is presented in Section~\ref{sect:res:stack:cos}. 

%In the stacking
%analyses we use the same algorithms to measure the blue/red flux ratio as in
%the individual cases.  

\subsection{Systemic redshift determination}\label{sect:zsys}

Where \zsys\ is not available it must be estimated from the \lya\ emission, for
which we again use algorithms from the LASD \citep{Runnholm.2020lasd}, which
are designed to provide a accurate average properties of a sample.  The
algorithms require an initial estimate for the redshift, for which we take the
estimate from \citet{Urrutia.2019} for the high-$z$ galaxies, and the measured
\zsys\ for the low-$z$ samples.  The algorithm then uses a decision tree and
works on the heuristic that \lya\ can be either single or double-peaked: it
first calls a peak identifier to determine the number of peaks, which allows
for N peaks, each of which must exceed SNR=5.  In the case of doubly peaked
\lya, the algorithm calculates the velocity of each peak and assigns \zsys\ to
the average value; in the case of a single peak it assumes the line corresponds
to a redshifted peak, determines its maximum, and then uses as `walker' in the
blueward wavelength direction and assigns \zsys\ to the velocity at which the
derivative of the flux density changes sign.  We note that the two methods
offer differing levels of precision, with the two-peak method giving an average
dispersion ($1~\sigma$) of 41~\kms, and the walking algorithm around 2.5 times
higher, where both sub-samples are selected to have clear double or single
peaks (44 and 38 galaxies, respectively).  Our double-peaked method gives a
sample dispersion that is marginally lower than that reported by
\citet{Verhamme.2018} for a similar method, and slightly worse in our
single-peak/walking algorithm method.  Systematic effects also differ somewhat
between the two branches of the decision tree: the average offset of the
double-peak algorithm is $-23$~\kms\ (around half the dispersion) and
$-66$~\kms\ for the walking algorithm, although this offset is also corrected
for on average (see below).  For the final redshift we ascribe the median value
computed over a Monte Carlo simulation of 100 realizations, in which the data
are randomized before the peak identification is performed: thus the decision
of which algorithm is used is also simulated.  This makes it difficult to
estimate the overall effects of the two algorithms on the final performance,
but note that in \citet{Runnholm.2020lasd} we show that the central 68~\% of
the distribution falls across a range of only 100~\kms.  That study also adopts
mixed redshift determination algorithms in the Monte Carlo simulation, includes
galaxies of somewhat lower signal-to-noise ratio than those selected here
(Section~\ref{sect:data:cos}), and is approximately consistent with the results
of both methods presented in \citet{Verhamme.2018}.

\begin{figure}
\includegraphics[width=8.5cm]{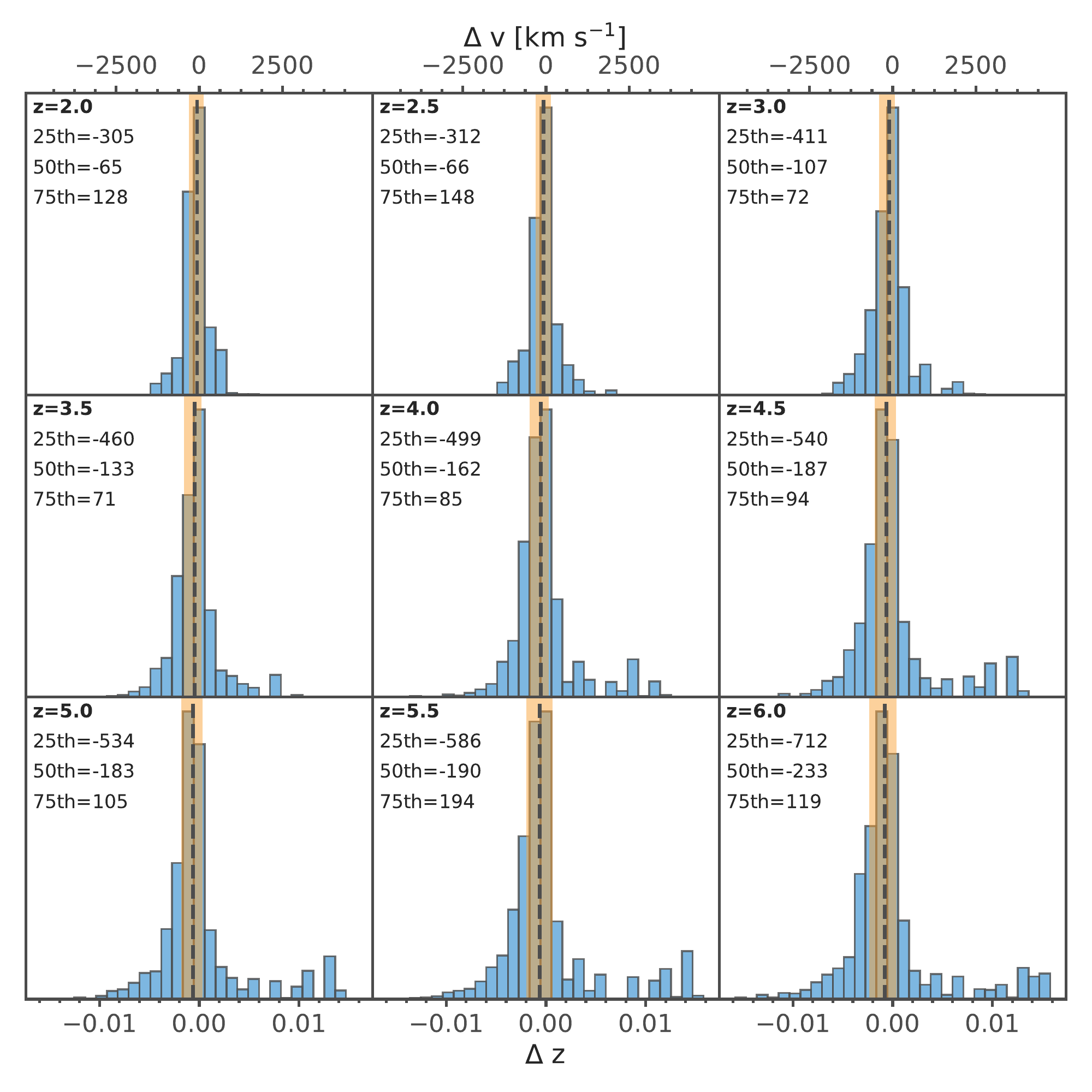}
\caption{The performance of the systemic error estimation algorithm.  The
discrepancy for each object in the Monte Carlo simulation is given as $\Delta
z$, which is defined as $z_\mathrm{LASD} - z_\mathrm{sys}$, where \zsys\ is
derived from optical emission lines.  The difference in velocities is given on
the upper axis.  The median value is shown with the dashed black line, and
interquartile range (IQR) in the orange region.  Median and IQR are also given
as inset text, as velocities in \kms.}
\label{fig:zsyshist}
\end{figure}

In this study we apply an additional correction based upon systematic offsets
observed within the sample at higher redshifts.  We adopt the same COS sample
as previously described and, using the systemic redshifts measured from optical
emission lines, we correct the spectra to restframe wavelengths and then
artificially redshift them to redshifts of 2.5 to 5.5 in steps of 0.5.  We then
run a library of random IGM realizations at each of these redshifts using the
method described in Section~\ref{sect:abssim} and attenuate each of the
redshifted spectra 50 times.  We run the $z$-detection algorithm on each galaxy
at each redshift, and examine the representative statistics of the output
compared to the known \zsys.  We show the results in Figure~\ref{fig:zsyshist}. 

The algorithm recovers \zsys\ with a median offset of $-65$~\kms\ at $z=2$,
which grows to $-233$~\kms\ by $z=6$.  This is a result of two phenomena:
firstly the blue peaks become systematically more suppressed with increasing
redshift, because the IGM absorption preferentially removes weaker blue peaks.
This causes the algorithm to more frequently switch from two-peak to one-peak
mode as redshift increases.  Moreover the thicker IGM as $z$ approaches 6
starts to influence the red peak.  Damping wings become visible in a minority
of simulations, but the blue wing of the red peak becomes systematically more
removed, shifting our $z$ estimate to higher positive velocities. 

\begin{figure}
\includegraphics[width=8.5cm]{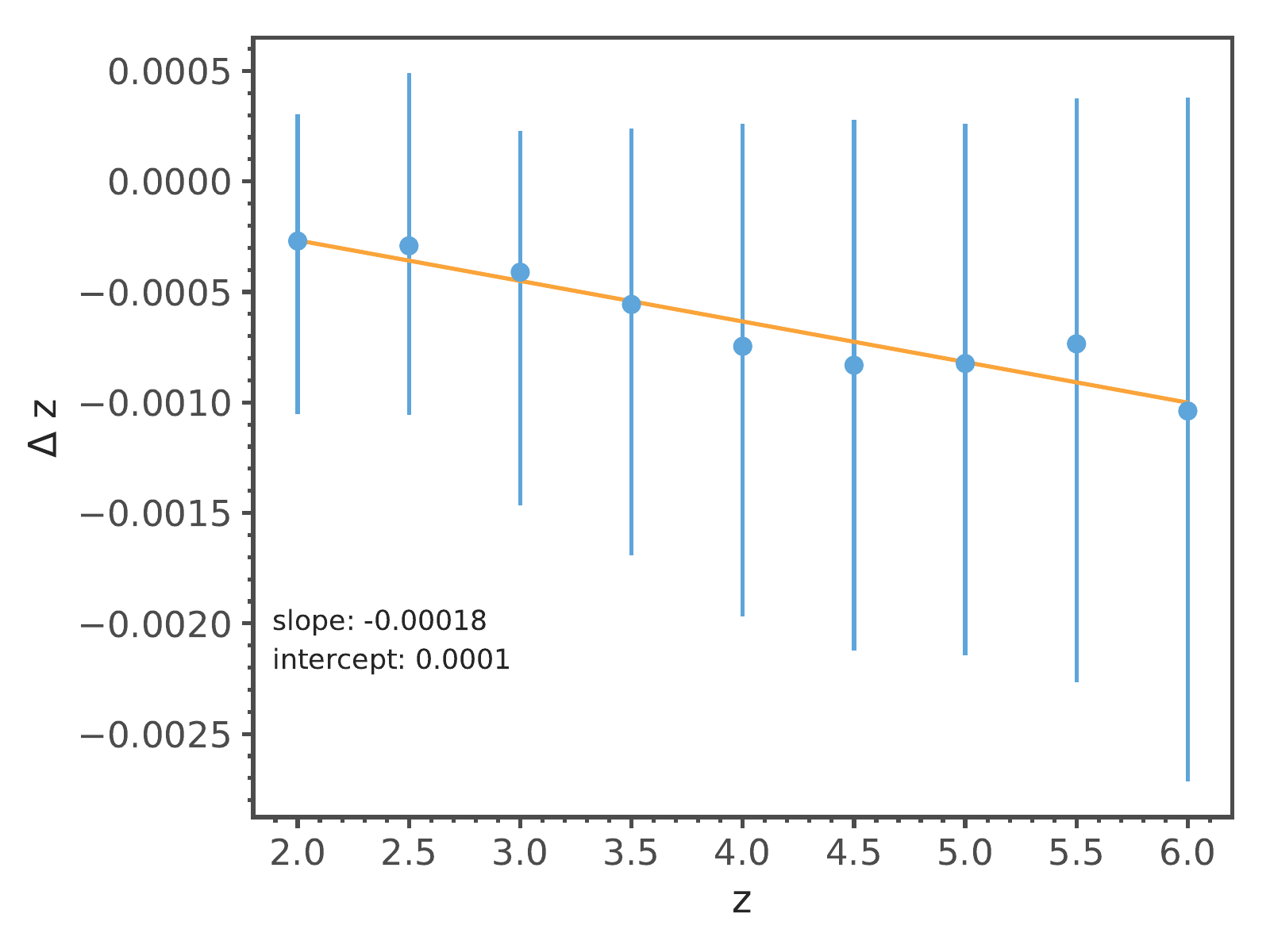}
\caption{The performance of the \zsys\ identification with redshift.  Redshift
is shown on the abscissa, while $\Delta z$ is shown on the ordinate-axis.  Each
point illustrates the median value from the simulated sample, and the error-bar
shows the IQR.  Our best fit is a linear function with the coefficients shown
in the Figure.}
\label{fig:zsysevol}
\end{figure}

This shift of the estimated \zsys\ with redshift is both expected and
predictable, and we show the redshift evolution of $\Delta z$ in
Figure~\ref{fig:zsysevol}. 

\subsection{Matching the spectral resolution}\label{sect:specres}

The comparative studies rely upon contrasting high-$z$ spectra for which the
spectral resolution is 85--160~\kms, with low-$z$ spectra that could in
principle have resolutions 5--6 times higher.  We first attempt to
approximately match the information contents of the spectra, although begin by
noting that (a.) the actual difference in resolutions will be much
smaller than would be inferred by simply contrasting the instrumental
specifications, and (b.) we cannot know the resolution of COS at \lya.  This
latter point arises because HST/COS has an aperture many times larger than its
focused point spread function, and the resolution consequently depends upon the
angular size of the source in the dispersion direction.  For \lya\ emission,
this in turn depends upon the size of the galaxy, the effects of \lya\
scattering, and the distance.  For example: LARS\,14
\citep{Hayes.2013,Hayes.2014} has half-light diameter in \lya\ of $\sim
3.6$~kpc (1.2~arcsec), which is larger than the unvignetted region of the COS
aperture; this scale length would reduce the effective resolution from $R\sim
15$,000 to below $R\lesssim 5000$.  \citet{Yang.2017size} report \lya\ full
width half maxima of 0.8~\arcsec\ for their sample of COS-observed galaxies
(most of which are included here), which is exactly the size of the unvignetted
region
\footnote{https://hst-docs.stsci.edu/cosihb/chapter-2-special-considerations-for-cycle-28/2-10-choosing-between-cos-and-stis}.
\citet[][whose observations are also subsumed into our sample]{Henry.2015} use
the NUV acquisition images and the line-spread-function (LSF) to estimates
spectral resolutions of $\sim 25-46$~\kms, but this applies only to the
continuum: assuming the linear size of \lya\ twice that of the UV on average
\citep{Hayes.2013}, these numbers would reduce to $\sim 50-100$~\kms, which now
overlaps with the MUSE range.   

For this study, we attempt to obtain a coarse match of spectral
resolutions, that will hold for the sample average.  For each galaxy we look up
$R$ at the wavelength of \lya\ from the figures given the Instrument Handbook,
and assume a \lya\ radius of 0.35~\arcsec\ (the median of
\citealt{Yang.2017size}).  We then compute the smoothing kernel that would be
required to match the spectra to $R = 3500$ by Pythagorean subtraction.  This
value is chosen to match the highest resolution possible with MUSE.  We finally
smooth the COS spectra with this kernel.  Again we note that this procedure
will not be perfect, but also that any more detailed method would still be
limited by the unknown light distribution of \lya\ which cannot be known
without an extensive imaging campaign.

\subsection{Simulating the intergalactic medium absorption}\label{sect:abssim}

We follow the method described in \citet{Inoue.2014}, which is based upon
statistics of absorption line systems observed in the spectra of quasars at
redshifts 3--6.  Inoue et al compile observed distribution functions for the
column density (\nhi) of absorbing systems from \lya\ forest (LAF) clouds to
damped \lya\ absorbing (DLA) systems, along with their evolution with redshift;
they then defined a set of power-laws to analytically describe the evolution of
the LAF, DLA, and Lyman limit systems (LLS).  The column density distribution
$dn/dN$ is a steep negative power law, and the redshift evolution $dn/dNdz$
evolves upwards with redshift.  With the parameterized PDFs and an assumed
target redshift, one can draw random sightlines for absorbing clouds and their
column densities.

Any sightline to $z>3$ will encounter a large number of \hI\ clouds that will
absorb radiation bluewards of \lya.  To apply this absorption,
\citet{Inoue.2014} adopt the Doppler parameter ($b$) distribution derived from
observations of LAF clouds \citep{Hui.1999,Janknecht.2006}, although given the
high oscillator strength of \lya\ this matters only for absorption at the very
lowest column densities.  With a sightline of absorbing clouds, each of which
has randomly drawn $z$, \nhi\ and $b$, one can compute the absorption across
the whole spectrum: Voigt profiles are used for discrete absorption in Lyman
series transitions, while photoionization bluewards of the Lyman would also be
needed for the transfer of ionizing radiation.  As we are only interested in
radiation just bluewards of \lya, only this transition is actually needed for
our purposes.  We implement the set of equations described in
\citet{Inoue.2014} so for any sightline we can randomly ascribe an absorption
spectrum, which gives IGM throughput as a function of wavelength.  In this
paper we perform extensive Monte Carlo simulations with this algorithm to
statistically study the effect of IGM absorbing clouds on the \lya\ profile.   

A recent relevant comparison of \lya\ emission and absorber statistics has been
made by \citet[][using the same MUSE-WIDE data as here]{Wisotzki.2018}.  They
rely mainly upon references included in \citet{Inoue.2014}, with the addition
of \citet{Crighton.2015}; these authors do revise the DLA incidence rate at
$z\sim 5$ somewhat compared to \citet{Prochaska.2009}, but the incidence rate
of proximate DLAs remains negligible compared to lower column density systems
that dominate the absorption in the \lya\ resonance.  Similar approaches have
recently been taken by \citet{Steidel.2018} using absorber statistics from
\citet{Rudie.2013}, but are optimized to study LyC absorption and do not appear
to treat the Doppler parameter independently.  The \citet{Rudie.2013}
statistics anyway appear to be consistent with those of \citet{OMeara.2013}
that are used by \citet{Inoue.2014}.

The statistics of these absorbing populations and their evolution are both
completely empirical and very well determined, and the great advantage of
absorption selection is that it suffers no redshift bias.  The numbers
presented by \citet{Inoue.2014} will represent a very realistic average
absorption in a spatially uncorrelated universe.  Note that this implies that
the PDFs of the absorbers are treated as being completely independent, and the
approach does not account for spatial correlation of matter in the universe.
This approach therefore contains no information on the CGM of the target
galaxies, or over-densities/clustering around them.  On the other hand,
galaxies identified by observation will be subject to CGM absorption and may
have nearby companion galaxies in the foreground.  Thus the regions probed by
galaxy surveys may be over-dense on average compared to these IGM prescriptions
(entirely randomized absorption).  Several authors
\citep{Santos.2004,Dijkstra.2007igm,Mesinger.2008,Iliev.2008,Laursen.2011,Gronke.2020}
have discussed the effect of infall and structure local to galaxies, that will
enhance the absorption of \lya\ because of the increased neutral column
crossing $\Delta z=0$.  One of the key results of our study is that
the excess absorption provided by the CGM is not necessary to explain the
evolving \lya\ profiles.  Since the absorption line statistics are implemented
randomly and drawn from Monte Carlo simulations, they provide a lower limit in
comparison to absorption in the case of a structured universe.  On the other
hand, such simulations also neglect the local enhancement of the ionizing
background that would be produced by both the observed galaxy and neighboring
systems.  These sources of ionizing radiation would act in the opposite
direction and suppress the local absorption.  While we acknowledge the possible
discrepancy between the average cosmic absorption and the enhanced opacity that
should occur close to galaxies, one of the key findings of this paper is that
\emph{there is no need for this additional absorption} (see
Section~\ref{sect:res:igmevol}).

\section{Basic Sample characterization }\label{sect:sample}

%The measurements we make in this paper were all made using algorithms developed
%for the \emph{Lyman alpha Spectral Database} (LASD; Runnholm et al 2020b) or
%from the measurements distributed and compiled by \citet{Urrutia.2019}.  Here
%we present a brief overview.  We also present a basic characterization in
%Section~\ref{sect:character}, and the first investigations of the line profile
%shapes can be found in in Section~\ref{sect:res:stack}.  

%\subsection{Basic sample characterization}\label{sect:character}

\begin{figure*}
\includegraphics[width=16cm]{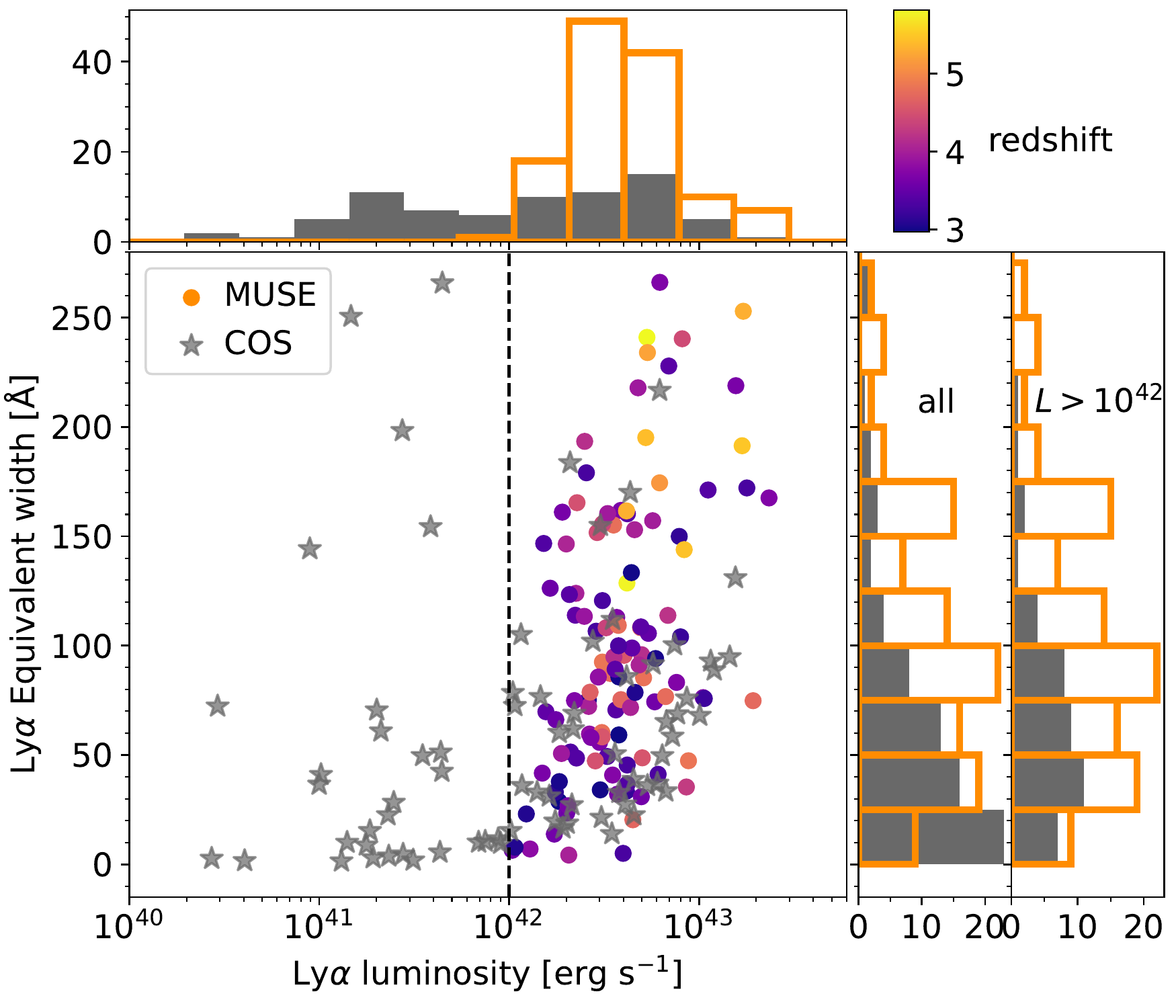}
\caption{The range of \llya\ and \ewlya\ spanned by the COS sample (grey stars)
and MUSE samples (colored circles).  The MUSE data are color-coded by redshift,
with the colorbar shown to the upper right.  A histogram showing the \llya\
distribution is shown above, and histograms showing \ewlya\ are shown to the
right: the first includes the total COS sample, while the second shows the COS
sample after retaining only galaxies more luminous than the cutoff of
\llya$>10^{42}$~\ergsec; this luminosity cutoff is illustrated by the dashed
black line.}
\label{fig:sample_compare}
\end{figure*}

We present the \lya\ luminosity and EW distributions of the two complete
samples in Figure~\ref{fig:sample_compare}.  The main panel shows the relation
between the two quantities, while the vertical histogram above shows the
distribution of \llya.  The two horizontal histograms to the right show the
total distribution of \ewlya\ in the inner panel, while in the outer panel the
COS sample has been restricted to \llya$\ge 10^{42}$~\ergsec; this cut is
designed to match the luminosities between the two samples, and is shown by the
vertical dashed line in the main figure.  The sample matching is discussed in
more detail below and in Section~\ref{sect:samplematch}. 

The range of EWs overlaps well between the sample, spanning values above $\sim
200$~\AA, even in the UV-selected low-$z$ sample.  The shapes of the EW
distribution exhibit similar declines towards higher values, but different
behavior in the lowest EW bins: being \lya\ selected, the MUSE sample is
under-represented at the lowest EWs (0--25~\AA) in this binning, while this is
the most populated bin the UV-selected COS sample.  The two samples are very
different in \llya: both samples extend to a comparably high luminosity
($\approx 10^{43}$~\ergsec), but the COS sample extends down to \llya~$\approx
3\times 10^{40}$~\ergsec\ while the MUSE data stop at \llya$=10^{42}$~\ergsec.
In the high-$z$ study that follows (Section~\ref{sect:samplematch} and
onwards) we retain only the low-$z$ galaxies with \llya$>10^{42}$~\ergsec, so
as to match \llya\ between the samples; this also has the effect of bringing
the EW distribution more into agreement (right histogram).  First, however, we
perform a more general study of the line profiles in the stacked spectra to
better understand the origin of the line profiles (blue bumps in particular)
and include all COS galaxies so as to preserve the dynamic range in the sample. 

%%%%%%%%%%%%%%%%%%%%%%%%%%%%%%%%%%%%%%%%%%%%%%%%
\begin{figure*}
\includegraphics[width=9cm]{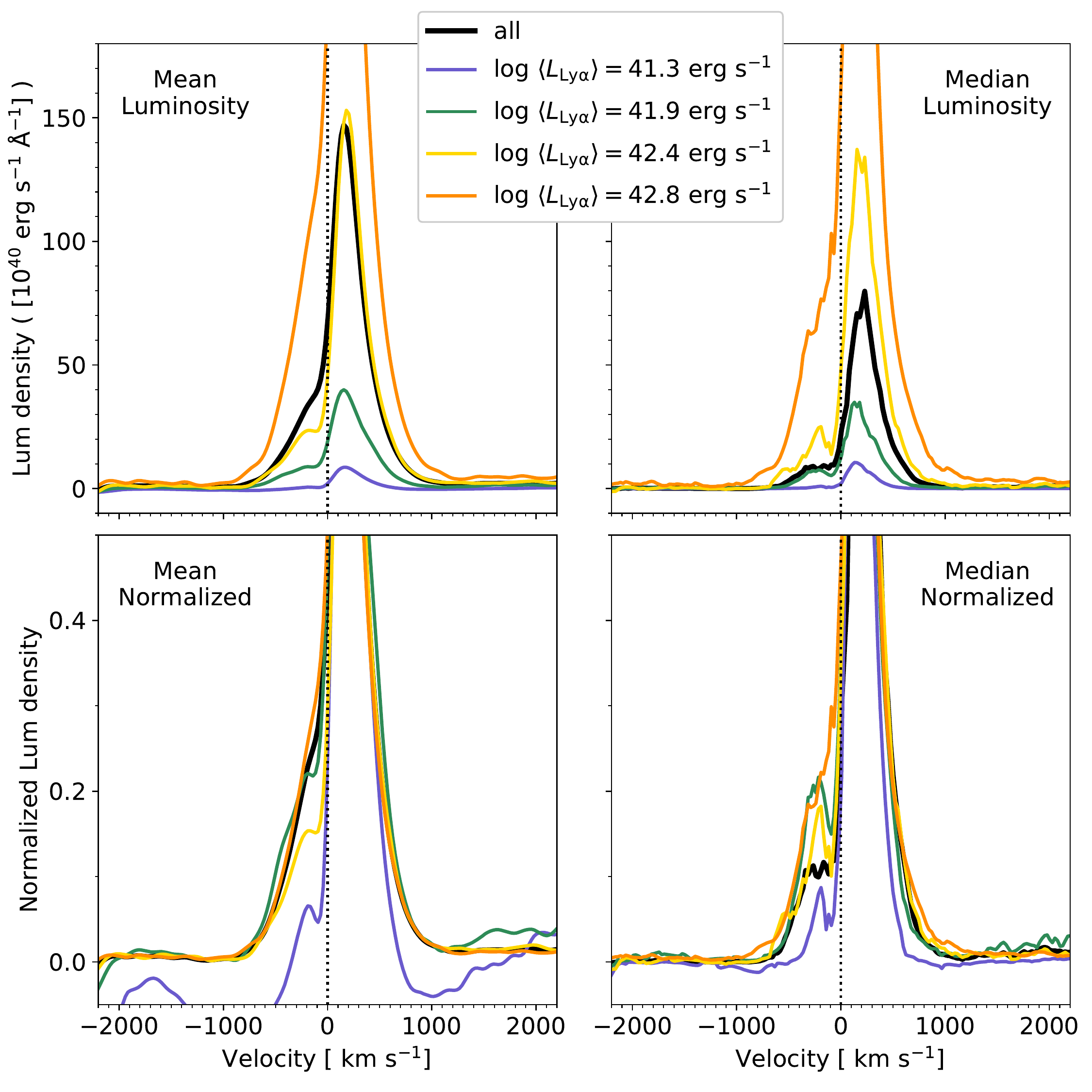}
\includegraphics[width=9cm]{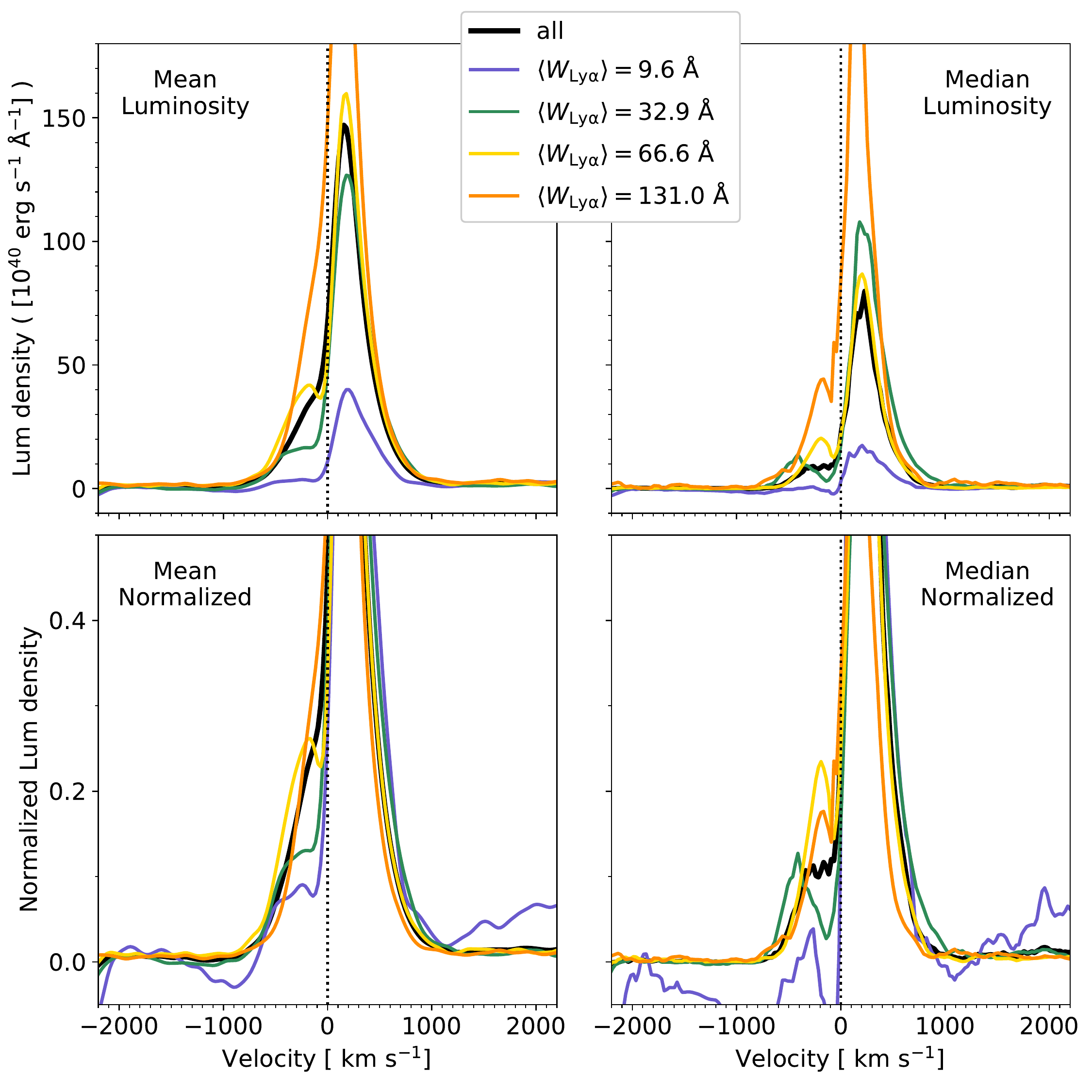}
\caption{Each panel shows a stacked spectrum from the full sample of galaxies
at $z<0.4$ observed with HST/COS.  We compare the spectral profiles as a
function of \llya\ (left $2\times2$ grid) and \ewlya\ (right $2\times2$ grid).
The sub-bins, divided by luminosity or EW are color coded with the values shown
in the legend, while the total stack of the whole sample is shown in black.  In
each block of four panels, the left pair shows the mean combination and the
right pair shows the median.  Upper panels show the spectrum in absolute units
of luminosity, while the lower panels show the spectra normalized by the
maximum luminosity density, which is always that of the red peak.  In the lower
panels we zoom in on the luminosity axis so as to emphasize the blue wings.  } 
\label{fig:cos_prop_stack}
\end{figure*}
%%%%%%%%%%%%%%%%%%%%%%%%%%%%%%%%%%%%%%%%%%%%%%%%

\section{Lyman alpha profiles: results from stacked
spectra}\label{sect:res:stack}

The bulk of the data analysis in this paper comes from differential comparison
of stacked spectra.  Because galaxies lie at different distances we perform the
stacking on spectra shifted into the restframe: we divide $(1+z)$ out from the
wavelength vectors, and multiply the same factor back in to the flux vectors
(always units of \ergseccmaa) so that the equivalent widths are preserved.  We
multiply all flux vectors by $4\pi d_\mathrm{L}^2$ to obtain luminosity density
vectors, with units erg~s$^{-1}$~\AA$^{-1}$.  Every spectrum is then resampled
onto the same velocity grid and co-added: we always record both the mean and
median spectrum.  We note that a well-known limitation of spectral stacking
analyses is that by reducing the spectrum to a single number at each frequency,
the variations within the sample are lost.  This dispersion is re-captured in
our case by resampling: when presenting stacked spectra we assess the variation
around each average profile by performing a bootstrap analysis of the spectra
contained within each bin. 

\subsection{HST/COS data at $z<0.4$}\label{sect:res:stack:cos}

We show the combined \lya\ profiles of the COS sample in
Figure~\ref{fig:cos_prop_stack}, where we examine how the profile shape varies
as a function of luminosity and EW.  It comes as no surprise that when dividing
out the sample by \llya, galaxies become more luminous (upper plots to the left
of Figure~\ref{fig:cos_prop_stack}).  We note, however, that there appears to
also be some evolution in the blue wing of the \lya\ profile in that it appears
to become disproportionately stronger in galaxies of higher \ewlya.  This
becomes much more obvious when the spectra are normalized by the maximum value
of the red peak prior to coaddition: in the lower panels to the left of
Figure~\ref{fig:cos_prop_stack} it is clear that as the luminosity decreases,
so does the relative contribution of the blue bump, which drops systematically
from the orange line (almost $10^{43}$~\ergsec) to the purple line (100 times
less luminous). 

We suspect this first result comes from variations in the column of blueshifted
absorbing material.  There is no absorption from the intergalactic medium at
this redshift, and outflows/galaxy winds result in a stronger absorption from
interstellar material on the blue side of \lya.  This becomes even more marked
in the lowest EW bin, where the mean combination reveals true blueshifted
absorption of the stellar continuum, and perhaps even a wing from damped
absorption extending onto the red side of \lya\ at a lower level.   At
intermediate luminosities, the wing contribution clearly increases from rough
10\%, up to 20--30~\% for the brightest galaxies. 

%%%%%%%%%%%%%%%%%%%%%%%%%%%%%%%%%%%%%%%%%%%%%%%%
\begin{figure*}
\includegraphics[width=18cm]{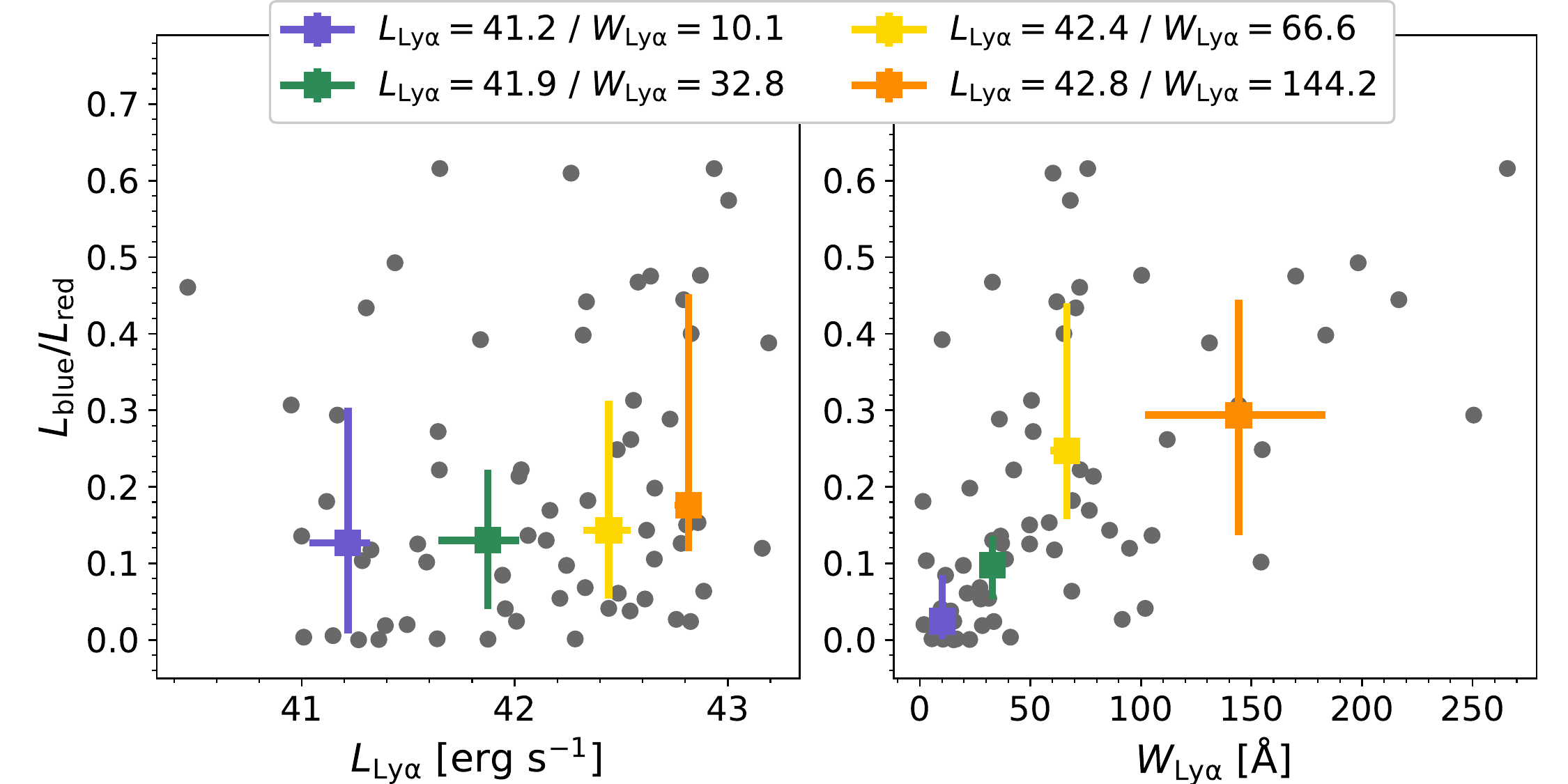}
\caption{The relative contribution of the blue \lya\ peak, expressed as the
\Lbluered\ ratio as a function of \llya\ and \ewlya.  It is similar to
Figure~\ref{fig:cos_prop_stack} but on a galaxy-by-galaxy basis.  The blue and
red \lya\ peaks are measured using the LASD software.  Colored points are
medians with interquartile ranges, and match with the bins in
Figure~\ref{fig:cos_prop_stack}.} 
\label{fig:cos_bluered_ew}
%u
\end{figure*}
%%%%%%%%%%%%%%%%%%%%%%%%%%%%%%%%%%%%%%%%%%%%%%%%

We examine this trend further in Figure~\ref{fig:cos_bluered_ew}, where in grey
we show how the blue/red flux fraction contrasts with \llya\ and \ewlya\ for
the individual galaxies.  Targeting positive blue-side emission only, we only
include here galaxies for which the flux density bluewards of line center
exceeds zero, and the majority of galaxies from the least luminous bin from
Figure~\ref{fig:cos_prop_stack} are removed.   \Lbluered\ shows significant
scatter at all luminosities and EWs.  Overlaid in color we show the median
values for each sub-sample, and encode the interquartile range (IQR) as the
shaded region.  IQR is a commonly used measurement of dispersion associated
with the median representative statistic, including data from the 25$^{th}$ to
75$^{th}$ percentiles; for a normal distribution, the IQR is $1.35\sigma$.
Attending to the left-most plot, \Lbluered\ appears close to invariant of
\llya\ at the low \llya\ end, but much of this effect comes from the removal of
absorbing galaxies and pure P\,Cygni-like systems from the bins.  In the last
bin, however, \Lbluered\ does jump by a factor of $\approx 2$ to 0.34, but
overall there is no correlation.  However, a tight correlation is seen between
\Lbluered\ and \ewlya\ (right figure), with Spearman's $\rho=0.60$,
corresponding to $p=7.6\times 10^{-8}$ off 68 data-points.  This result is
highly significant, and indeed the points show that the lowest EW galaxies have
\Lbluered\ at the percent-level, which rises dramatically to $\approx 30$~\% at
\ewlya\ of $\approx 100$~\AA.  This result quantitatively echoes those
demonstrated in \citet{Henry.2015} and \citet{Yang.2017prof}, whose \lya\ EWs
are among the highest observed and whose profiles also include the highest
fraction of double peaks.  Note, however, that the Henry et al. and Yang et al.
observations are subsumed into our sample, and these results are not entirely
independent. 

%%%%%%%%%%%%%%%%%%%%%%%%%%%%%%%%%%%%%%%%%%%%%%%%
\begin{figure}
\includegraphics[width=8.5cm]{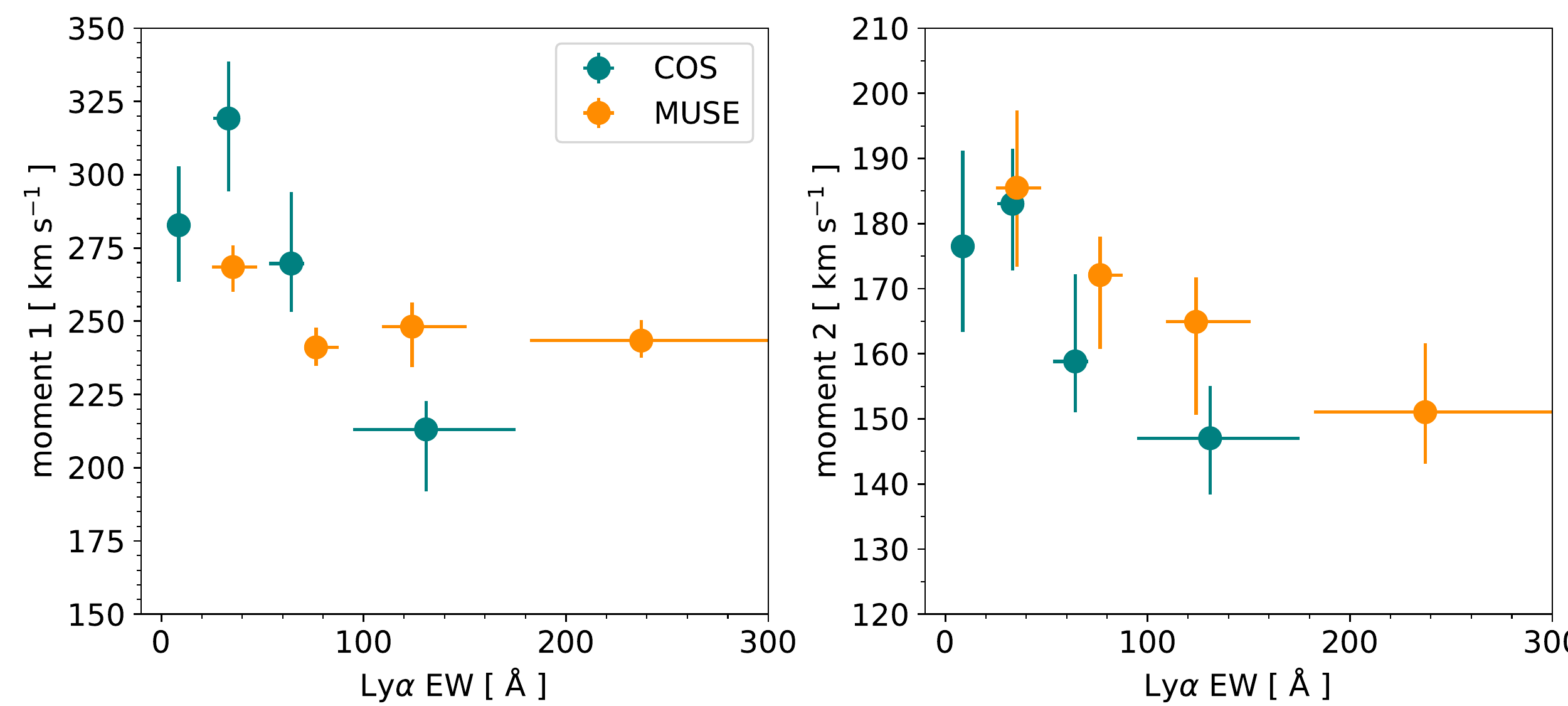}
\caption{The evolution of the first and second moments of the red peak as a
function of \ewlya.  The complete range of luminosities of the COS sample are
used and shown in green; the MUSE sample, which is more luminous but spans less
dynamic range, is shown in orange.  Median values and interquartile ranges are
shown.} 
\label{fig:mom12_vs_ew}
%u
\end{figure}
%%%%%%%%%%%%%%%%%%%%%%%%%%%%%%%%%%%%%%%%%%%%%%%%

We also note that galaxies with higher \ewlya\ also show smaller systematic
offsets in velocity (moment 1) and narrower red lines (moment 2), as shown in
the normalized median profiles to the lower right left in
Figure~\ref{fig:cos_prop_stack}.  The second moments of each bin decrease
monotonically from 185~\kms\ at \ewlya~$\approx 30$~\AA\ (shown by both MUSE
and COS) to about about 150~\kms\ at \ewlya~$\approx 230$~\AA\ (probed only by
MUSE), as shown in the right panel of Figure~\ref{fig:mom12_vs_ew}.  We expect
this is due to an effect of \hI\ column density in which larger columns result in
more \lya\ scattering, which reduces the total emitted \ewlya\ as the escape
probability decreases.  This additional scattering further results in broader
\lya\ lines that are systematically more redshifted.

\subsection{VLT/MUSE data at $z>2.9$}\label{sect:res:stack:muse}

%%%%%%%%%%%%%%%%%%%%%%%%%%%%%%%%%%%%%%%%%%%%%%%%%
\begin{figure*}
\includegraphics[width=9cm]{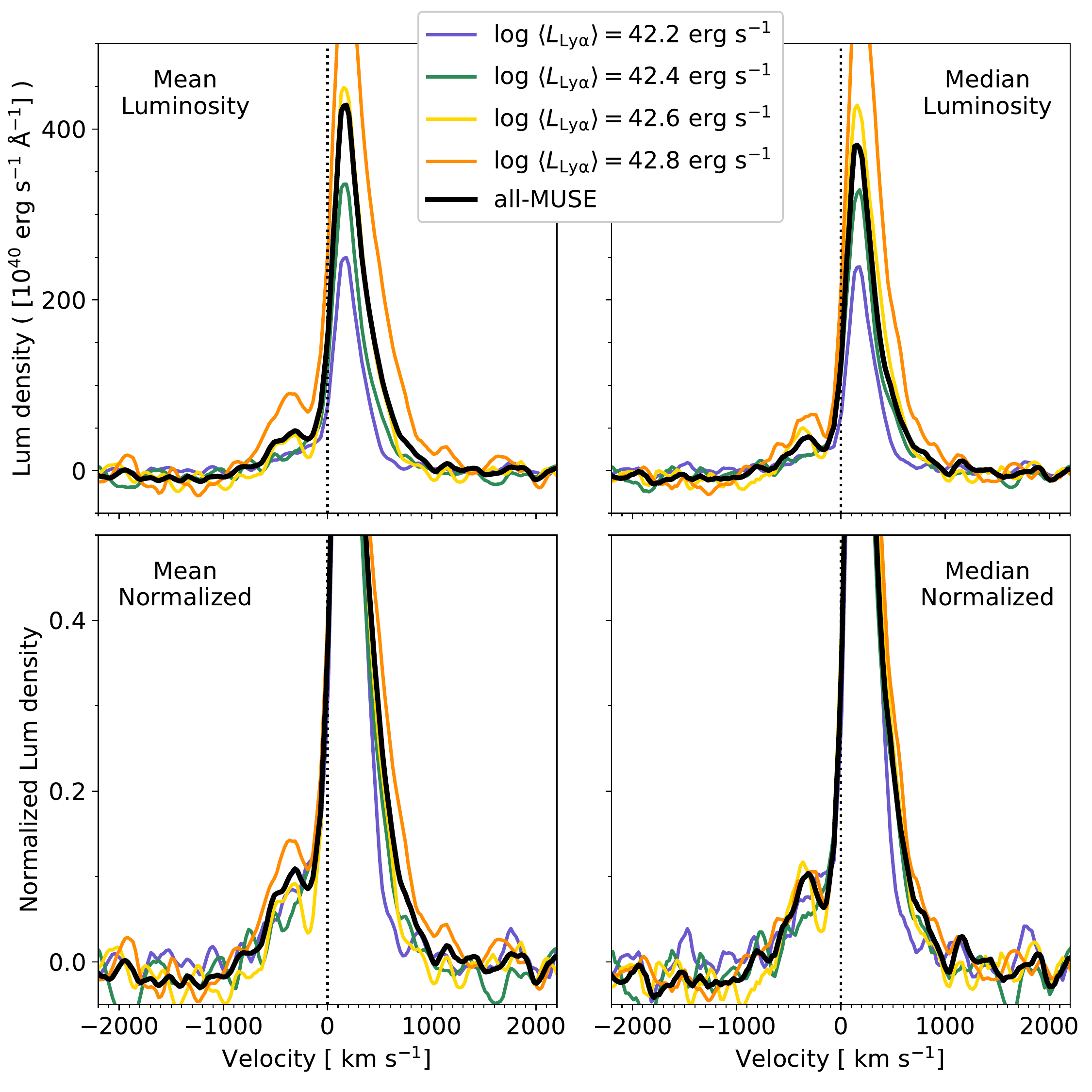}
\includegraphics[width=9cm]{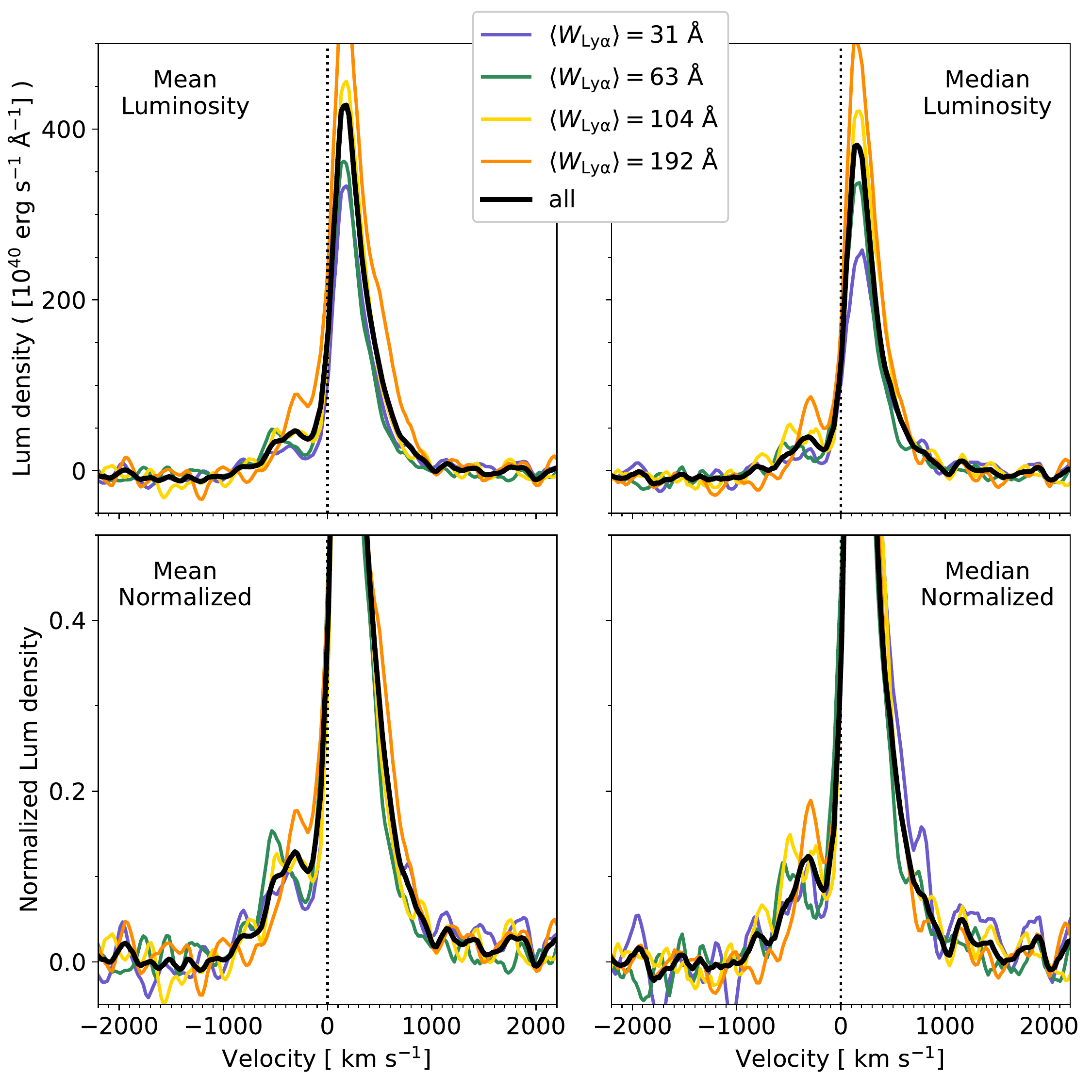}
\caption{Same as Figure~\ref{fig:cos_prop_stack}, but for the MUSE-WIDE sample,
selecting only galaxies at $2.9<z<4$.} 
\label{fig:mw_prop_stack}
\end{figure*}
%%%%%%%%%%%%%%%%%%%%%%%%%%%%%%%%%%%%%%%%%%%%%%%%%

Figure~\ref{fig:mw_prop_stack} shows the stacked spectra obtained with
VLT/MUSE.  At these redshifts the IGM may have significant influence over the
\lya\ profile, especially in the blue side, so we temporarily restrict our
analysis to $z<4$ galaxies, so as to compromise between studying less absorbed
galaxies while still maintaining a significant sample.  Blue wings are also
visible in the MUSE spectra, and examining all the lower panels of
Figure~\ref{fig:mw_prop_stack} it is clear that some emission is visible to
velocities as high as $\Delta v\approx -700$~\kms.  However unlike the
lower-$z$ COS spectra, this blue emission is visible in every bin of luminosity
and EW.  While the lines are not necessarily in agreement within the noise,
relative intensity in the blue bumps does not show a serial decline as it does
at lower-$z$: e.g. the green line clearly out-lies the others in the EW
comparison, but this is likely attributable to random processes.  

In order to explore this lack of a trend further, we perform a bootstrap study
using the low-$z$ objects (no IGM absorption), and the IGM simulations
described in Section~\ref{sect:abssim}.  We first replace each galaxy in the
$2.9 < z < 4$ MUSE sample with the low-$z$ galaxy that has the closest \llya.
We then randomly draw a sightline through the IGM to the MUSE redshift,
attenuate the COS galaxy, and bin and stack this random sample.  We examine the
frequency with which a statistically significant ($p<0.05$) trend is recovered.
With samples the size of this MUSE subset ($N=74$) the correlation is almost
never reproduced: with just a few percent of realizations showing significant
correlations.  However when we artificially increase the bootstrap simulation
to include $\approx 1000$ objects (including galaxies more than once but with
different random IGM sightlines) the trend does re-emerge, with the \Lbluered\
ratio reduced by a factor of 2 compared to $z=0$.  This is exactly what would
be expected for large samples, as the stochastic effect of the IGM is smoothed
out over a large number of realizations. 

In principle this method could be used to calculate the necessary sample size
required in order to statistically study these trends at high-redshift, but
many other factors will hold influence over the result.  For example the
precise recovery will also depend upon the signal-to-noise ratio of both the
high-$z$ data and the low-$z$ data we use to simulate it, and the result will
also be redshift dependent.  Nevertheless we proffer the advice to use low-$z$
data from the extensive and growing HST archives to realistically simulate the
recovery of high-$z$ \lya\ spectral profiles in large samples.

Interestingly, the same trend in the width and offset of the red peak is
observed in these higher-$z$ systems, as discussed for $z=0$ galaxies in
Section~\ref{sect:res:stack:cos}.  The normalized profiles of
Figure~\ref{fig:mw_prop_stack} (lower left) both show the lower-\ewlya\
galaxies to have both broader and less shifted red profiles; same is shown
quantitatively in the right panel of Figure~\ref{fig:mom12_vs_ew}.
\citet{Hashimoto.2015} show the red-peak velocity shift to be much higher in
LBGs than in LAEs using optical line emission such as \halpha\ and [\oIII], and
similar effects are probably at work here.  They attributed this effect to
radiation transfer effects where the LBG column density is higher, using RT
modeling in spherical shells. \citet{Cassata.2020} also show similar effects
using the [\cII] 158~\micron\ line: they show weak correlations of \lya\
redshift with UV magnitude, stellar mass and SFR.  These go in the direction
that would support the hypothesis that mass drives the trend, but none of
Cassata et al's trends reach statistical significance, possibly because of
dynamic range.

\subsection{Comparison of matched samples}\label{sect:samplematch}

The COS sample from low-$z$ is needed for an estimation of the intrinsic \lya\
line profile.  However, to produce such an intrinsic, IGM-free template for the
high-$z$ galaxies, we would ideally match the samples in their SFRs, masses,
compactness, dust optical depths, ionizing intensities, etc.  Such sample
matching is far in the future and the huge majority of this information is
unavailable at $z>2.9$; obtaining it will require extensive surveys with the
James Webb Space Telescope (JWST). We instead make the most basic compromise,
and take only COS galaxies that overlap with the MUSE sample in the two key
observables, of \llya\ and \ewlya.  The distribution of these quantities is
shown in Figure~\ref{fig:sample_compare}, where the full COS sample extends to
30 times lower luminosities than the faintest MUSE-WIDE system.   This is
certainly to be expected when contrasting distant \lya-selected galaxies with
nearby UV-selected systems.  

We impose a cut in the \lya\ luminosities at $10^{42}$~\ergsec, as shown by the
dashed line in Figure~\ref{fig:sample_compare}.  We do not aim to accurately
reproduce the distributions in \ewlya\ and \llya, but note a number of
similarities.  Firstly the upper envelope of the luminosity histogram is
well-matched between the samples.  In the \ewlya\ histograms, the MUSE-WIDE
survey is skewed towards somewhat higher values, with a median \ewlya\ of
78~\AA\ compared with 61~\AA\ for the low-$z$ sample.  The \lya\ blue/red flux
ratio depends strongly upon \ewlya, and increases towards higher EWs as also
shown in Figure~\ref{fig:cos_bluered_ew}, and this effect could introduce a
bias in the direction that high-$z$ galaxies have larger intrinsic blue/red
ratios.  While we do not see the same behavior in the high-$z$ galaxies
directly, we showed this to result from the stochastic IGM absorption in
Section~\ref{sect:res:stack:muse}.  On the other hand, the median EWs of the
two samples remains close (much smaller than the interquartile range), and
while acknowledging the EW difference could introduce a bias we expect it to be
a minor one.  

We finally note that the COS aperture (radius of 1\farcs25) samples only a
projected distance of 5~kpc at $z=0.25$, and should part of the line profile be
built from extended emission then that would obviously not be captured by the
low-$z$ observation.   Indeed \citet{Leclercq.2020} do find differences in the
velocity centroids of \lya\ profiles extracted from the core and halo regions,
with the implication that asymmetric profiles are built from differences
between the core and halo \citep[see also][]{Erb.2018}.  However we note that
these conclusions from the deep MUSE data were drawn from a sub-sample in which
the halo flux fraction was high by construction, skewing the interpretations in
the direction of the more luminous and extended LAEs.

In the coming Section on redshift evolution, we use the COS sample to estimate
the intrinsic and absorbed \lya\ line profile from high-$z$.  From now on,
however, we take only low-$z$ galaxies from the luminosity matched sub-sample
($N=45$).

\section{The Co-Evolution of Lyman alpha profiles and the intergalactic
medium}\label{sect:res:zevol}

In this Section we present our main results concerning the evolution of the
\lya\ emission line with redshift and how we can explain this in terms of the
co-evolution of galaxies and the IGM.  

\subsection{The Redshift Evolution of the \lya\
Profile}\label{sect:res:profevol}

\begin{figure*}
\includegraphics[width=18cm]{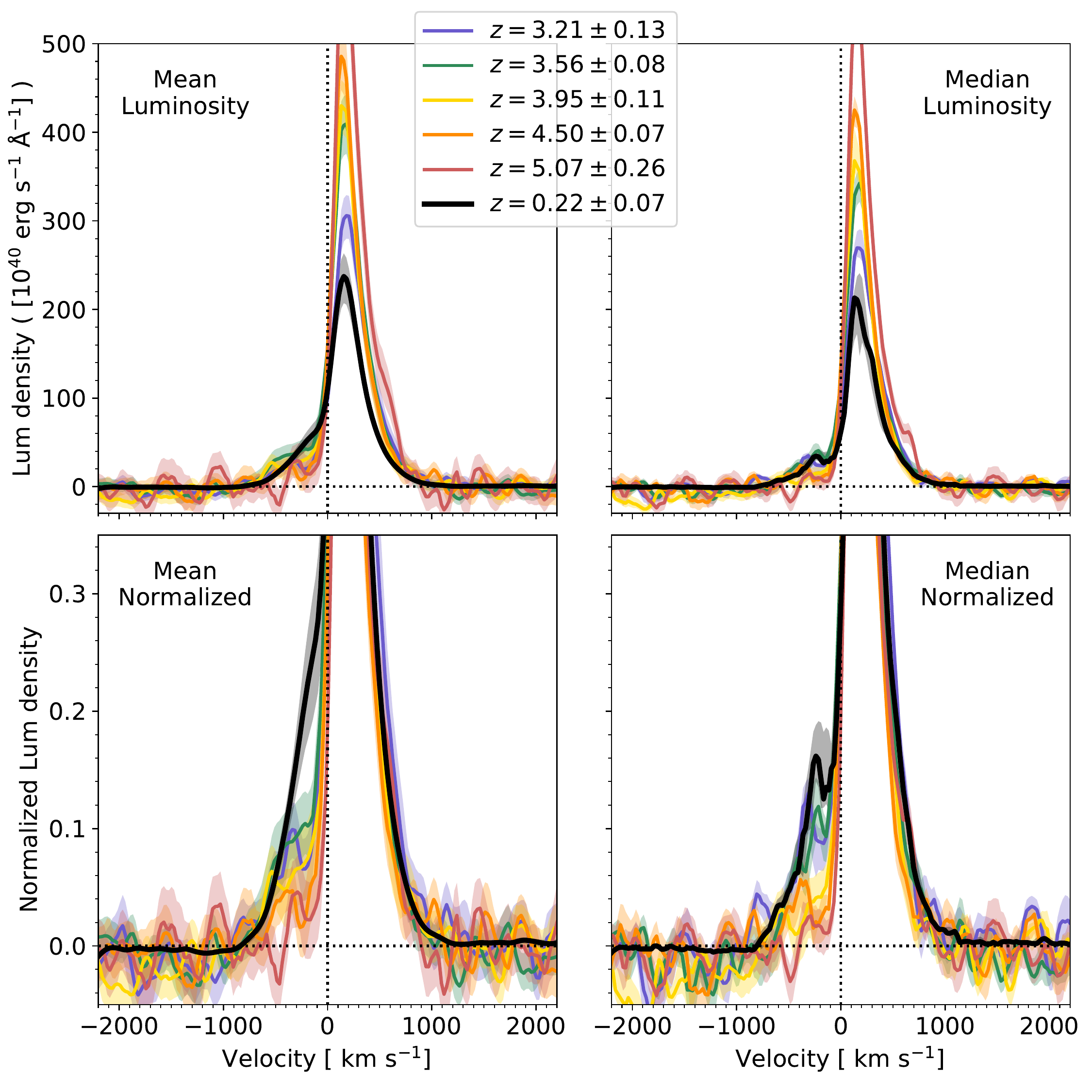}
\caption{The evolution of the \lya\ profile with redshift.  The individual
panels are the same as in Figures~\ref{fig:cos_prop_stack} and
\ref{fig:mw_prop_stack}, but the sample is now binned by redshift; each bin
includes 45 or 46 galaxies.  The MUSE-WIDE sub-samples are shown in colors and
the redshifts are displayed in the caption.  The low-$z$ sample observed with
HST is shown in black.  Error regions are shaded, and represent the
interquartile range estimated from bootstrap resampling.  The evolution of the
blue wing, decreasing with increasing $z$, is obvious.} 
\label{fig:mw_redshift_stack}
\end{figure*}

We now proceed to study the redshift evolution of the \lya\ profile, using
differential stacking comparisons similar to those already shown.  We bin the
sample in redshift into five sub-samples, and show the results in
Figure~\ref{fig:mw_redshift_stack} with sample medians and half-quartile ranges
shown in the caption.  The high-$z$ galaxies from the MUSE-WIDE survey are
always shown in color, and the $z\approx 0.25$ sample, with no IGM, in black.
For the low-$z$ sample, we adopt precisely the same strategy as in
Section~\ref{sect:res:stack:cos}, except switch the estimate of \zsys\ from the
measured value to that estimated from the \lya-line using the LASD algorithms.
This ensures consistency in the treatment of the samples.  Despite changing the
\zsys\ estimator, the COS profiles remain very similar, and the black line in
Figure~\ref{fig:mw_redshift_stack} is close to the average of the yellow and
orange lines in the lower left panels of Figure~\ref{fig:cos_prop_stack}.
Especially for the more luminous galaxies, the errors on the recovered systemic
redshifts become comparable with the resolution of the spectrograph. 

At each redshift we resample the galaxy bin using bootstrap techniques.  We
randomly re-draw a sample of the same size and recompute the mean and median;
iterating this over 1000 realizations we compute the variations of each stacked
subsample about the true mean and median.  For mean profiles we show the
central 68$^{th}$ percentiles, and for the median profiles we show the
interquartile range. 

Several effects are clear from Figure~\ref{fig:mw_redshift_stack}.  Among the
most obvious, visible in the upper two panels there is a trend among the MUSE
wide galaxies to become more luminous with increasing $z$.  This is the
well-known \citet{Malmquist.1922} bias, that manifests as the most luminous
systems being preferentially detected.  In this instance it results from the
luminosity distance increasing with $(1+z)$ faster than the sensitivity of MUSE
decays with $(1+z)\times \lambda_\mathrm{Ly\alpha}$: even though the \lya\
luminosity function is close to constant across this redshift range
\citep[e.g.][]{Herenz.2019}, there is a upwards evolution the sample
luminosities because only the most luminous subset of the respective galaxy
population is detected at highest redshifts.  

Blue wings are visible in the profiles.  This is most abundantly clear in the
COS profiles from low-$z$, but sub-dominant blue wings are visible in the
un-normalized profiles (upper).  However their intensity evolution goes in the
opposite direction from the evolution in the red peaks: the highest intensity
blue peaks are found among the $z\approx 3-4$ subsamples, and the \Lbluered\
ratios decrease with $z$.  This is more noticeable when we normalize the
profiles, which we show in the lower two panels.  Here we zoom in to visualize
the blue peaks: attending mainly to the median stack (lower right) the
$z\approx 3-3.5$ sub stacks show normalized intensity of 10~\%, which drops
further to about 5~\% by $z\approx 4$ and almost to zero at $z=4.5$ and up.  

In light of the Malmquist bias discussed above, a natural question becomes
whether this evolution in the line profiles may result from selection effects
or evolution with luminosity.  This phenomenon is seen within the IGM-free COS
sample (Figure~\ref{fig:cos_prop_stack}) but not in the MUSE sample
(Figure~\ref{fig:mw_prop_stack}), and we showed with simulations
(Section~\ref{sect:res:stack:muse}) that it would be impossible to detect for
the given sample size and spectral quality even at $z=3-4$.  Importantly, the
median luminosity evolution due to the Malmquist bias is only a factor of 2
between $z=3$ and 5.5, which is significantly smaller than the difference
between the bin median in Figure~\ref{fig:cos_prop_stack} where the trend is
seen.  We conclude that the blue peak evolution presented in
Figure~\ref{fig:mw_redshift_stack} is not attributable to selection effects of
this kind.

Another potential source of bias could be the errors on the estimated \zsys:
these errors increase with redshift (Figure~\ref{fig:zsysevol}), and higher
errors at $z\sim 5-6$ could artificially suppress the blue peaks.  We have
tested the impact of these increasing \zsys\ errors by adding an additional
error term to the redshifts of the lower-$z$ galaxies: we compute this from the
fit in Figure~\ref{fig:zsysevol}, and derive the magnitude of the random
deviate needed to match the breadth of the distribution at $z\approx 6$.  We
find no significant effect on the profile evolution.

The COS spectrum from $z\approx 0.25$ shows a \Lbluered\ ratio of $\approx
16$~\% in the median stack, and higher still in the mean.  Mean stacks,
however, will always be dominated by the more luminous systems at a given
velocity, and result in line profiles that are on average less representative
of a given galaxy.  We observe a reduction in the normalized blue peak
intensity of about $\approx 60$~\% to $z=3$; from a simple by-eye analysis we
would estimate the an IGM opacity of \tauigmlya$\approx 0.45$ from
Figure~\ref{fig:mw_redshift_stack}.  In the coming Section we perform a more
detailed study of this.

\subsection{The Impact of the IGM} \label{sect:res:igmevol}

\begin{figure*}
\includegraphics[width=18cm]{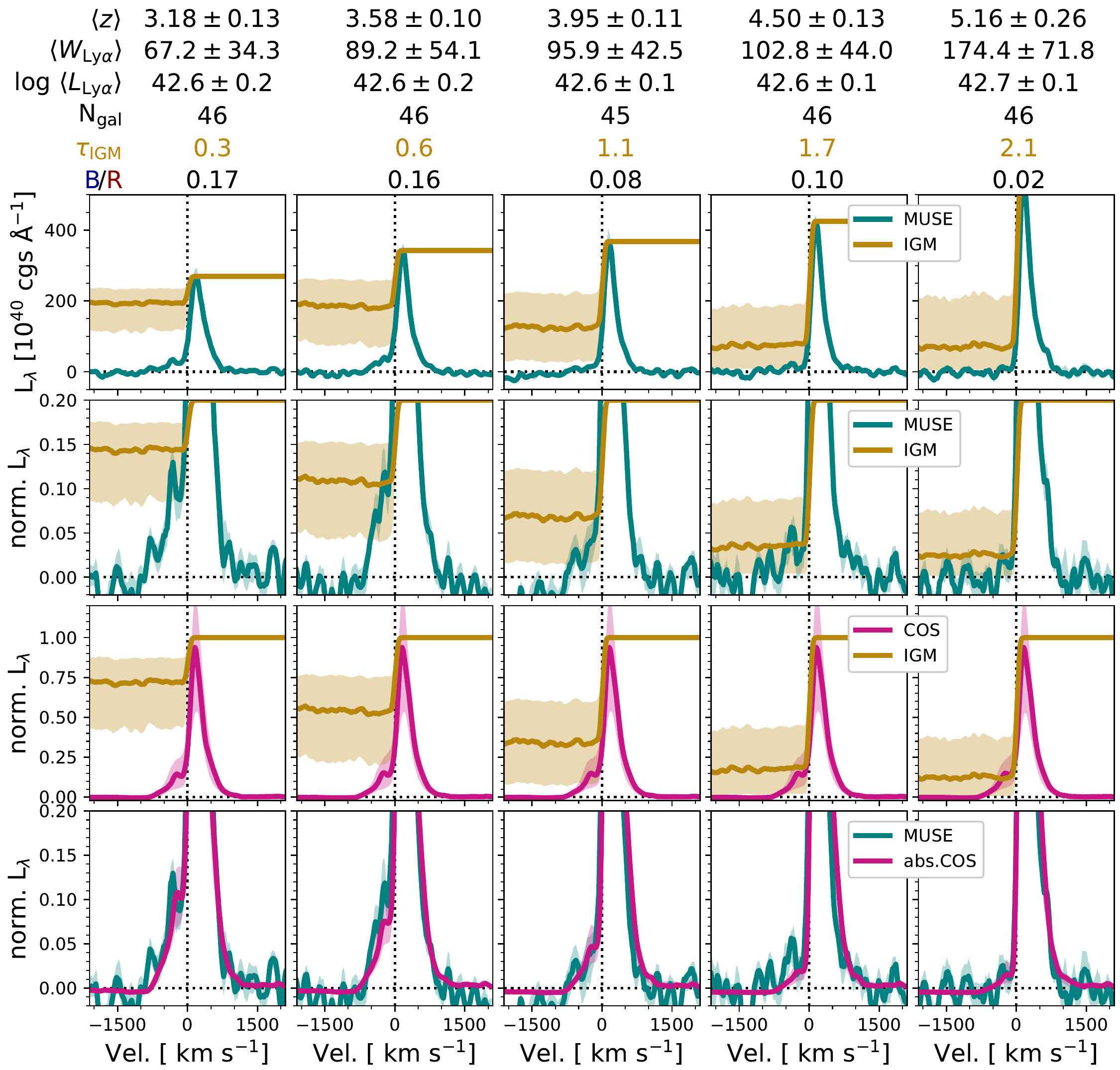}
\caption{Simulating IGM absorption of LAEs at various redshifts.  Each column
shows a different redshift bin: the median $z$, \ewlya, \llya\, are listed
above, together with the half-quartile range of each quantity, and the number
of galaxies in each bin.  The top rows show the median stack of the \lya\
luminosity density in absolute units (erg~s$^{-1}$~\AA$^{-1}$) in green, with
the median IGM absorption overlaid, scaled so that 1 matches the red peak of
the \lya.  Shaded regions represent the interquartile range.  The second row
shows the same \lya\ profiles normalized to an intensity of 1 at the peak, and
zoomed in around the wings of the line: IGM is again shown on arbitrary scale
of $0.15\times$. The third row shows the intrinsic spectrum in pink,
which is simulated from an equal number of galaxies drawn from the luminosity-
and EW-matched samples from $z\approx 0$.  This spectrum is always normalized
to a peak intensity of 1, and the shaded region shows the quartile range
derived from bootstrap simulations (see Section~\ref{sect:res:igmevol} for
details).  The IGM transmission is also shown in the third row}, now
scaled to the absolute throughput.  The average optical depth, \tauigmlya, is
computed in the $\lambda_\mathrm{rest}=1200-1210$~\AA\ range, and is shown at
the top of each column.  The lowest row shows the profile of the COS spectrum,
absorbed by the IGM for each redshift bin, together with the stacked observed
MUSE spectra for the same redshift.  The overlap between the two lines is often
striking.
\label{fig:allz_absorb}
\end{figure*}

The COS spectra are not susceptible to intergalactic \lya\ absorption: the IGM
optical depth at $\lambda=1200-1210$~\AA\ is \tauigmlya~$\lesssim
10^{-6}$.  Thus if the \lya\ line profile observed in the COS sample is
representative of the profile emergent from the CGM at redshifts beyond 3, this
empirical spectrum will serve as a template.  We restate that we selected the
low-$z$ sample to match the observables of the MUSE-WIDE sample in both \lya\
luminosity and EW (Figure~\ref{fig:sample_compare}), which argues in favor of
this utilization.  

We present our findings on the redshift evolution of the \lya\ profiles,
together with simulations of the evolving IGM opacity, in
Figure~\ref{fig:allz_absorb}.  Each column shows data and simulations for a
different redshift bin, and is headed with the following: redshift, \ewlya,
\llya, the number of galaxies in the bin, \tauigmlya, and the \Lbluered\ ratio
from the MUSE-WIDE stacks.  The left column shows the results from the lowest
redshift, centered at $z=3.18$.  The average \lya\ profile from the stack of 46
high-$z$ galaxies is shown in the top row (this is the same as the red
spectrum in Figure~\ref{fig:mw_redshift_stack}).  At this redshift the blue
wing is very clear, and is visible out to velocities of $-700$~\kms;
this is especially clear in the second row, where we zoom in to show
the blueshifted emission in detail. The IGM throughput ($I/I_0$) is shown in
yellow in various rows: in the top two panels it is arbitrarily scaled, while
in the third row it is presented in absolute throughput: in the first column,
the average transmission just bluewards of \lya\ is $\approx 0.75$,
corresponding to the \tauigmlya~$= 0.3$ quoted in the heading.

The third and fourth rows show the simulated \lya\ profile (pink lines) at each
redshift, in the unabsorbed and absorbed cases, respectively.  We simulate the
emergent \lya\ profile using a randomly drawn sub-sample of $z\approx 0$
spectra, taking $N$ COS observations, where $N$ is the number of high-$z$
galaxies in the bin (46 in this example).  For this we consider only the
luminosity and EW-matched subsamples (Section~\ref{sect:samplematch} and
Figure~\ref{fig:sample_compare}).  Assuming no evolution in the intrinsic \lya\
profiles, we first normalize and median-combine the $N$ low-$z$ spectra, using
the same routines as for the MUSE data.  We finally run a 1000-realization
Monte Carlo simulation to build the distribution of expected profiles around
this median (IQR is shown as the shaded region).  This unabsorbed spectrum is
shown in the third row of Figure~\ref{fig:allz_absorb}.

For each galaxy discussed above, and in every realization of the Monte Carlo
simulation, we also add the effect of the IGM by artificially absorbing each
selected COS spectrum.  Prior to normalization, we multiply each spectrum by an
absorption spectrum, randomly generated for a single sightline through the IGM
to the redshift of the MUSE spectrum (see Section~\ref{sect:abssim} for
details).  We then proceed with the normalization, co-addition, and
distribution sampling in precisely the same way as for the unabsorbed spectra
in the previous paragraph.  The resulting median spectrum, together with its
quartile range, is shown by the pink line in the bottom row of
Figure~\ref{fig:allz_absorb}; the average high-$z$ spectrum is again shown in
green. Attending again to the lower left panel ($z=3.18$), the agreement
between the simulated spectrum and the observed spectrum at the same redshift
is striking.  The normalized flux goes to zero in the blue wing at velocity
offsets of $\lesssim -700$~\kms\ in both spectra; immediately bluewards of
line-center they differ by $\approx 20$~\% in normalized intensity, with the
simulated absorption from the IGM marginally over-predicting the observation
from $z\sim 3$.  

It is possible that this slight deficit in \lya\ flux at $\Delta v \gtrsim
-300$~\kms\ is due to enhanced absorption from structure/galaxies adjacent to
the target LAEs.  The IGM simulations follow the prescription of
\cite{Inoue.2014}, which is constructed using the observed distribution of
\hI-absorbing systems identified along the sightlines to luminous quasars.
Each absorber is drawn at random from a probability distribution function, and
is therefore independent of all other absorbing systems and the target galaxy
itself; it neglects the galaxy-galaxy spatial correlation and the fact that
these LAEs should occupy somewhat over-dense regions which will enhance the
absorption within the volume bound from the Hubble expansion by gravity.
Precisely this has been addressed in various numerical methods
\citep{Santos.2004,Dijkstra.2007igm,Mesinger.2008,Iliev.2008,Laursen.2011,Gronke.2020},
who have all used cosmological hydrodynamical simulations to assess the
distribution and velocity field of material within 10~Mpc of galaxies.  In
fact, such simulations have already been used to correct the \lya\ EW in
spectroscopic studies of high-$z$ galaxies \citep{Pahl.2020}.  While the
precise results will depend somewhat upon the details of the simulation and
mass ranges probed by the observation, the broad picture is clear: at $z
\approx 3-3.5$ sample there is a sharp excess of \lya\ absorption at velocities
$\approx -50$ to $-100$~\kms\ from line-center that increases the \lya\
absorption by a factor of two to three times.  This excess of absorption
returns to the baseline value (random absorption systems) at velocity offsets
exceeding $\approx 300$~\kms, and it seems entirely plausible that this
phenomenon could explain the deficit of blue-side flux close to line-center at
the lower redshift end of the MUSE sample.  However, we note while this excess
absorption due to circumgalactic material is visible for all halo masses at
$z\gtrsim 3$ in the simulations, it is not required to explain the stacked
spectra studied here. 

We now proceed to examine the next redshift bin, centered at $z=3.55$.  The
MUSE-WIDE galaxies at this redshift completely match with those of the previous
bin in \llya; and the median EWs are consistent within 8\AA\ ($\approx 25$~\%
the half-quartile range; values quoted above the figure rows).  By this
redshift the IGM optical depth has increased by a factor of two (0.3 to 0.6)
and provides significantly more absorption on average.  When we use this IGM
absorption profile to artificially attenuate the COS spectrum we bring the COS
and MUSE blue components into tight agreement at all velocities between $-1000$
and 0~\kms.  Indeed the absorbed COS spectrum completely overlies the MUSE
stack, with no need for excess absorption at any velocity; it is difficult to
tell which stack is which. 

The IGM opacity increases systematically, with \tauigmlya\ going from 1.1, to
1.8, to 2.3 at redshifts $z=3.9$, 4.5, and 5.4, respectively.  If we take the
typical $L_\mathrm{blue}/L_\mathrm{red}$ ratios observed at low-$z$ as
representative ($\sim 0.35$; Figure~\ref{fig:cos_bluered_ew}), then we would
expect the blue bumps to reach approximately 0.1, 0.06, and 0.03 in the
absorbed spectra that are normalized to the red peak.  This is precisely what
is shown in columns 3, 4, and 5 of Figure~\ref{fig:allz_absorb}: probable blue
bumps remain but if real their significance is low, and their normalized
intensity is reduced to the level of a few per-cent.  At none of these higher
redshifts do we need an additional component of absorption at low blueshifted
velocities, that we attributed to structure near to the galaxies or cosmic
infall \citep[e.g.][]{Dijkstra.2007igm,Laursen.2011}.  

\section{Discussion and Implications}\label{sect:discussion}

\subsection{Radiation Transfer Simulations in Realistic Environments}

In recent years a number of theoretical efforts have been devoted to predicting
the \lya\ output of galaxies from RT models, using two complementary
approaches.  Researchers have used idealized model galaxies described by a
number of single parameters (gas column density, outflow velocity, etc) to
examine how \lya\ profiles vary with these basic properties
\citep[e.g.][]{Verhamme.2006,Schaerer.2011,Gronke.2016}: the main criticism
levelled at these studies is that realistic galaxies are not well described by
such simplistic models, so the community instead turns to simulated galaxies to
produce the manifold for RT simulations
\citep[e.g.][]{Tasitsiomi.2006,Laursen.2007,Faucher-Giguere.2010,Zheng.2010,Barnes.2011,Trebitsch.2016,Smith.2019}.
These models are not without criticism either, and are often cited as having
insufficient spatial resolution, particularly at CGM scales, to capture the
representative geometry for the RT.  RT in these settings almost invariably
over-produces \lya\ emission on the blue side of line center, at a level that
is only very rarely found by observation.  The proponents frequently appeal to
intergalactic absorption to remove the blue component, but our results show
that this appeal is unlikely to be realistic.  The uncorrelated \lya\ absorbing
systems do not reach sufficient opacity to universally suppress the blue
emission seen in starburst galaxies, and we find no need within this large
dataset for an excess opacity near galaxies.  However, we also note that LAEs
such as these do reside in lower-mass halos, and there still may be room for
the effect to be more prominent on sightlines towards higher halo mass.

\subsection{Is there still space for excess opacity? }

The absence of enhanced \lya\ within a few hundred \kms\ of the galaxies is
worthy of further discussion.  One explanation could be that with structure
building up hierarchically over time, we need to wait until redshifts $\sim 3$
for significant \emph{neutral} gas to affect the \lya\ line to an extent
greater than the uncorrelated cosmic average.  One may speculate that the
galaxies studied here are less massive than those studied in the simulations
referred to above, and the amplitude of their over-density/infall compared to
the cosmic mean will be lower.  However with the typical range of stellar
masses in the mock galaxies being $\approx 10^7 - 10^{10}$~\msun\ this seems
unlikely on average as LAEs of this luminosity probably have $M_\mathrm{stell}
\approx 10^8-10^9$~\msun.  Alternatively, the gas may be more ionized than
thought, which would imply the simulated opacity is overestimated.  This might
be due to a lower clumping factor of cold, neutral material (e.g., because of a
lack of spatial resolution; \citealt{vanDeVoort.2019,Hummels.2019}) or
additional ionization sources.  We may point towards photoionization of
low-mass companion structures by the enhanced UV background close to early
galaxies, where the excess UV results from numerous galaxies that are too faint
to be detected.   Such galaxies may, however, be visible in the ultra-deep MUSE
pointings \citep[e.g.][]{Bacon.2017}.  

We also question whether this absence of excess blue-side absorption could be
the result of a conspiracy of parameters, where the intrinsic \Lbluered\ does
increase with $z$, but is suppressed by absorption on small scales.  Indeed the
Malmquist bias results in an upwards evolution of \llya\ with $z$
(Figure~\ref{fig:mw_redshift_stack}, upper) which may be expected to increase
the intrinsic \Lbluered\ (Figures~\ref{fig:cos_prop_stack} and
\ref{fig:cos_bluered_ew}).  However we note that the trends that are revealed
in the COS sample are manifested over a dynamic range of almost 2~dex, whereas
the overall evolution introduced by the Malmquist bias is just a factor of 2
between $z=3$ and 5.5.  In fact, attending to the sample medians shown at the
top of Figure~\ref{fig:allz_absorb}, there is very little evolution in total
(observed) \llya.  So while we cannot rule out a conspiracy of parameters, we
believe the dynamic range, coupled with the relatively small Malmquist bias,
are insufficient to produce an effect of the required magnitude. 

\subsection{Less blueshifted \lya, but more \lya\ overall}\label{sect:cosmicevol}

We note also that this downwards evolution of the blue contribution, and
clearly increasing impact of the IGM, occurs against the backdrop of an overall
\emph{increasing} \lya\ output from the galaxy population.  This is shown by
numerous lines of evidence including a \lya\ fraction among LBGs that increases
significantly over the same redshift range we probe here
(\citealt{Stark.2010,Cassata.2015,deBarros.2017}, although see also
\citealt{Caruana.2018}), an increase in the volume-averaged \lya\ escape
fraction of galaxies \citep{Hayes.2011evol,Dijkstra.2013,Konno.2016,Wold.2017},
and the increasing agreement of the UV LFs of LAEs and LBGs \citep{Ouchi.2008}.
In these first papers we could only speculate about the actual influence the
IGM has on the \lya\ emission: because the \lya\ output \emph{increases}  with
$z$, we concluded the IGM could not act to strongly suppress \lya.  The new
evidence presented here shows that this is not the case, and if the
COS-measured \Lbluered\ of $\approx 0.35$ continues to hold at high-$z$ (or
even grow), then the IGM removes an additional quarter of the total \lya\
(integrated over the whole line) at $z=5.5$ compared to 3.  The implication is
that the \lya\ emitted from galaxies must increase by a corresponding amount
over the same redshift range in order to counteract this effect.  Such excess
emission may be attributed to the decreasing dust content, stellar metallicity,
or average age of galaxies with redshift, implying they must evolve even faster
than previously claimed \citep[e.g.][]{Hayes.2011evol}.

\subsection{Implications of the evolving ISM and ionizing output of galaxies}

The blue/red flux ratio is often invoked as a tracer of the
interstellar medium of galaxies, especially regarding the \hI\ column
\citep[e.g.][]{Erb.2016} and ionizing output \citep[e.g.][]{Verhamme.2015}.
The absence of strong evolution in the intrinsic \Lbluered, therefore, raises
the question of whether the ionizing escape fraction (\fesclyc) is constant
between $z=0$ and $z=6$.  While we do not have the data necessary to answer
this question directly, it is likely that \fesclyc\ does not change
dramatically within the samples analyzed here.  However these samples comprise
a different fraction of the galaxy population at each redshift.

\citet{Steidel.2018} studied the LyC output of a large sample of UV-selected
galaxies at $z\sim 3$, and determined that the highest escape fractions are
found among galaxies with higher \ewlya.  Moreover, these same higher \ewlya\
galaxies also show more \lya\ emission close to $\Delta v=0$ than
weaker/non-emitting galaxies, which is consistent with our own results here at
$z\lesssim 0.4$ (Section~\ref{sect:res:stack:cos}).  It is likely that the
\lya\ selection of our sample, combined with the somewhat fainter UV continuum
magnitudes would place our galaxies towards the upper end of the \fesclyc\
distribution, at approximately \fesclyc~$\sim 20$~\%.  The same \fesclyc\
conclusion would also hold in the $z\gtrsim 5$ sample, but note that LAEs
comprise a larger fraction of the total galaxy population at this higher
redshift \citep[e.g.][and arguments in
Section~\ref{sect:cosmicevol}]{Ouchi.2008,Stark.2010}.  If anything, our result
of non-evolving line profiles would lend support to a picture in which the
volume-averaged ionizing output of galaxies evolves in line with the \lya\
fraction and/or the evolving \lya\ luminosity function.

Here we also recall that the average \fesclyc\ required to reionize the
universe must be on this order \citep{Robertson.2015}; while it may be somewhat
lower \citep[e.g.][]{Finkelstein.2019} it cannot be much higher as it would
violet the `photon-starved' nature of reionization \citep[e.g.][]{Bolton.2007}.
By the same token, the average \fesclyc\ at the lower redshift end of our MUSE
sample ($z\sim 3$) must occupy the same range so as not to over-heat the \lya\
forest \citep{Becker.2013}. 

A claim of an average \fesclyc\ of $\sim 20$~\% at $z\sim 0$ would indeed be
suspicious, as LyC emission is rarely reported in the more general starbursting
galaxy population \citep{Leitet.2013}.  It is only in the last few years have
LyC detections have systematically been made \citep[e.g.][]{Izotov.2018}; while
the \fesclyc\ estimates are again comparable (10--20~\% on average but with
large dispersion) these measurements derive from samples of specially selected,
rare compact starbursts with highly ionized ISM.  We again caution that the
fraction of strong \lya-emitting galaxies at these redshifts is an even smaller
fraction of the population than at $z\sim 3$ \citep[e.g.][]{Cowie.2010}.

\section{Outlook and limitations}\label{sect:outlook}

In Section~\ref{sect:sample} we showed that there is real and strong evolution
in the relative contribution of the blue wing of the \lya\ emission with
redshift, and in Section~\ref{sect:res:igmevol} that we are able to fully
reproduce this with IGM absorption.  The main assumptions were that the
intrinsic \lya\ spectrum is invariant with redshift and can be described by the
profile of low-$z$ sample observed with HST/COS, and that the IGM is spatially
uncorrelated and does not require an enhancement from companion
galaxies/proximate Lyman limit systems.  Furthermore the remarkable agreement
between the \lya\ profiles at all $z$, supports the utility of our improved
heuristic \zsys\ estimation from analysis of the \lya\ line profile
\citep{Runnholm.2020lasd}, despite an evolution in the IGM opacity (\tauigmlya)
of more than two optical depths.  This naturally raises the question of to what
extent the observed evolution of the \lya\ profile can be used to infer the IGM
opacity at even higher redshifts, and into the epoch reionization (EoR).

As mentioned in Section~\ref{sect:intro} strong blueshifted emission peaks have
also been found at $z=5.8-6.6$.  Some prime examples include Aerith B at
$z=5.8$ \citep{Bosman.2020}, NEPLA4 at $z=6.5$ \citep{Songaila.2018} and COLA1
at $z=6.6$ \citep{Hu.2016,Matthee.2018}, which all show spectacular \lya\
profiles, with dominant red lines but inarguable blue peaks.
\citet{Songaila.2018} suggest that such spectral features may be present in the
profiles of less luminous galaxies in significant numbers of up to 1/3.  This
is quite unexpected in light of the results shown in
Figures~\ref{fig:mw_redshift_stack} and \ref{fig:allz_absorb}, but two key
points must be born in mind when co-interpreting the datasets.  Firstly our
analysis is based upon stacking which acts as a lossy compression that removes
all information concerning the dispersion within the sample.  If blue bumps are
present at $z>5$ in the galaxies from the \citet{Urrutia.2019} data release, as
a significant minority, they will not be represented in the median stacking
(right columns) and will be hard to detect in the averaging.  Secondly the
$z>5.7$ blue-bump objects mentioned above are all significantly more luminous
than MUSE-WIDE galaxies: typical log~(\llya/(\ergsec)) is 43.5, while our
$z\approx5.2$ subsample has a median luminosity that is 0.8 dex lower.  

As we are likely unable to identify blue bumps in the MUSE-WIDE data, this
makes comparison to the more luminous systems difficult.  For a fixed \lya\
escape fraction, the `ordinary' systems studied here would themselves be almost
ten times less ionizing than the luminous blue-bump galaxies.  Nevertheless,
the overwhelming majority of the ionizing photons required to ionize a bubble
with radius at least 1~Mpc come from sources other than the galaxy itself, even
for more luminous galaxies \citep{Bagley.2017}, we should be wary of
over-interpreting spectra of single galaxies.   Even in the case of very large
\hII\ regions, the presence/absence of blue peaks/wings may have more to do
with the galaxy's position along the line-of-sight with respect to the bubble
edges, and the details of its ISM, than with the IGM state. 

One limitation of our study is that our galaxies are \lya-selected.  At the
higher ends of mid-$z$ ($z\sim 4-5.5$) this may remain a representative sample
of the galaxy population: the average \lya\ output of galaxies continually
increases \citep{Stark.2010,Hayes.2011evol}.  However all these \lya-quantities
begin to drop after $z\sim 6.5$ \citep{Schenker.2012}, which raises doubts
about the utility of LAE-based probes at higher redshifts.  At these redshifts
the \lya-based studies may provide representative statistics about the geometry
of \hII\ bubbles, but not about the regions between, which must still be probed
with (most likely) UV-selected/Lyman break galaxies. 

In the coming years, Subaru Prime Focus Spectrograph (PFS) will embark upon
very large area \lya\ surveys (14 deg$^2$ under the Subaru Strategic Program)
targeting UV-selected galaxies over what is likely to be the complete range of
bubble distributions in the EoR.  PFS can target \lya\ at $\lesssim 6.3$ at
$R\approx 5000$ with the mid-resolution red spectrograph, and at $\gtrsim 6.7$
at $R\approx 4300$ with the NIR spectrograph; the gap between the two redshifts
can be filled by the low-resolution red spectrograph at $R\approx 3000$, which
is comparable to MUSE at these redshifts.  An obvious study therefore becomes
the behavior of the \lya\ profile as a function of local galaxy environment, in
which the blue contribution to the total \lya\ can be used to constrain the
dispersion on \tauigm\ at fixed redshift.  This paper, and
Figure~\ref{fig:allz_absorb} in particular, shows that this goal can be reached
at the expected resolving powers of PFS.  
%Indeed the same goal cannot
%efficiently be reached the VLT/MUSE because of the small survey volume at the
%highest redshifts.  

The natural drawback of this fiber-fed approach, of course, is that it demands
photometric pre-selection, and will not have the possibility to identify new
LAEs within its own observation.  The SSP will target LBGs and LAEs from SSP
imaging surveys GOLDRUSH and SILVERRUSH, and as a consequence the \lya-emitter
components will be limited to the narrowband depths, which currently fall
around log~(\llya/(\ergsec))\,=\,43.0 for SILVERRUSH \citep{Konno.2018}.  Thus
the survey will deliver huge numbers of spectra with luminosities close to
those of COLA1 and NEPLA4, and LBGs from GOLDRUSH that are more massive still,
it remains unclear how well the survey will sample the mass range of more
`normal' star-forming LAEs at similar redshift.  Thus the observational
necessity remains for deep IFU studies with instruments like MUSE, that deliver
100\% spectroscopic completeness with the contrast and depth that can only be
provided by spectroscopic surveys.

\section{Summary and Conclusions}\label{sect:conclusions}

We have undertaken an extensive study of the \lya\ emission line profile from
star forming galaxies at various cosmic epochs, using some of the largest
samples of available spectra obtained with HST/COS at low-$z$ and VLT/MUSE at
$z=2.9$ to 6.6.  Our main findings can be summarized as:

\begin{itemize}
\item We see distinct and predictable variations in the shape of the \lya\
profile in the low-$z$ sample. The blue peak gets stronger with increasing
\ewlya.  We believe this trend results from atomic gas column being the
dominant factor in regulating \lya\ output, and as the galaxies almost
certainly have outflows the blue peak is more significantly suppressed than the
red. 
\item This trend is not observed within the MUSE sample at $z=2.9-4$ ($N=74$
galaxies).  Using Monte Carlo bootstrap simulations of the IGM opacity and
applying them to the low-$z$ sample, we show that the stochastic nature of IGM
opacity to \lya\ is entirely sufficient to remove trends of this amplitude. 
\item We find the red component of the \lya\ profile becomes broader and
increasingly redshifted at lower \lya\ equivalent widths.  We attribute these
effects to the covering of atomic material.  More scattering increases the
absorption of \lya\ and preferentially in the blue wing while, at the same
time, the larger column density requires longer frequency excursions to escape,
resulting in broader, more redshifted lines.
\item Using basic \lya\ observables of \llya\ and \ewlya, we identify a
\lya-matched sample of galaxies at $z\lesssim 0.4$.  The COS sample shows a
significantly higher average blue peak than the $z\gtrsim 3$ galaxies, with
roughly twice the intensity in the blue bump.  The blue-bump amplitude of the
high-$z$ sample decreases systematically from $z=3$ and upwards, becoming
undetectable by $z\approx 5.5$.
\item The evolution of the average observed \lya\ profile with redshift is very
well described by adopting the average profile from low-$z$ and absorbing it
with the current best estimates for the IGM opacity as a function of redshift.
This IGM prescription does not account for line-of-sight correlation in the
absorbers, and assumes they are random in redshift -- we do not find evidence
for additional absorption of \lya\ resulting from small scale structure and
peculiar motions/infall in adjacent to galaxies. 
\end{itemize}

\section*{Acknowledgements} 
We are grateful to the core members of the MUSE-WIDE team -- Lutz Wisotzki,
Tanya Urrutia, and E. Christian Herenz -- for making the extracted spectra
available and enabling this project.  We also thank Lutz Wisotzki and Tanya
Urrutia for useful discussions on the MUSE data, repeatability, and pointing
out a bug in the original manuscript.  We also thank Peter Laursen for
providing useful insights into the CGM and evolution of the IGM, and Anne
Verhamme for helping us understand the systematics involved in systemic
redshift estimation.  M.H. acknowledges the support of the Swedish Research
Council, Vetenskapsr{\aa}det and the Swedish National Space Agency (SNSA), and
is Fellow of the Knut and Alice Wallenberg Foundation.  M.G. was supported by
NASA through the NASA Hubble Fellowship grant HST-HF2-51409 and acknowledges
support from HST grants HST-GO-15643.017-A, HST-AR-15039.003-A, and XSEDE grant
TG-AST180036.

\bibliography{a.bib}{}
\bibliographystyle{aasjournal}

\end{document}